\documentclass[fleqn,usenatbib]{mnras}
\usepackage{amsmath,amssymb,physics,bm,esint}
\usepackage{physics,commath}
\usepackage{orcidlink}
\usepackage{color}
\usepackage{listings}
\usepackage{graphicx}
\usepackage{textcomp}
\usepackage{natbib}
\usepackage{soul}
\usepackage{comment}
\usepackage{placeins} 
\usepackage{xlmacros}
\usepackage{tikz}
\usepackage{lineno}

\usetikzlibrary{shapes.geometric, arrows}
\tikzstyle{data} = [rectangle, rounded corners, minimum width=2cm, minimum height=1cm,text centered, draw=black, fill=red!30, align=center]
\tikzstyle{filter} = [rectangle, rounded corners, minimum width=2cm, minimum height=1cm,text centered, draw=black, fill=blue!30, align=center]
\tikzstyle{obs} = [rectangle, rounded corners, minimum width=2cm, minimum height=1cm,text centered, draw=black, fill=green!30, align=center]
\tikzstyle{res} = [rectangle, rounded corners, minimum width=2cm, minimum height=1cm,text centered, draw=black, fill=orange!30, align=center]
\tikzstyle{arrow} = [thick,->,>=stealth]

\renewcommand{\vec}{\bm}
\newcommand{\dtilde}[1]{\tilde{\tilde{#1}}}

\DeclareRobustCommand{\VAN}[3]{#2}
\let\VANthebibliography\thebibliography
\def\thebibliography{\DeclareRobustCommand{\VAN}[3]{##3}\VANthebibliography}

\title{Analytical Noise Bias Correction for Precise Weak Lensing Shear Inference}
\author[X. Li et al.]{
Xiangchong Li$^{1, 2}$\thanks{xli6@bnl.gov}\orcidlink{0000-0003-2880-5102},
Rachel Mandelbaum$^{1}$\orcidlink{0000-0003-2271-1527},
The LSST Dark Energy Science Collaboration \\
$^{1}$ Department of Physics, McWilliams Center for Cosmology, Carnegie Mellon
University, Pittsburgh, PA 15213, USA \\
$^{2}$ Brookhaven National Laboratory, Bldg 510, Upton, New York 11973, USA
}

\date{Received August 12, 2024; accepted Month ??, 2024}
\pubyear{2024}

\begin{document}

\label{firstpage}
\pagerange{\pageref{firstpage}--\pageref{lastpage}}
\maketitle

\begin{abstract}
    Noise bias is a significant source of systematic error in weak
    gravitational lensing measurements that must be corrected to satisfy the
    stringent standards of modern imaging surveys in the era of precision
    cosmology. This paper reviews the analytical noise bias correction method
    and provides analytical derivations demonstrating that we can recover shear
    to its second order using the ``renoising'' noise bias correction approach
    introduced by \metacal{}. We implement this analytical noise bias
    correction within the \anacal{} shear estimation framework and propose
    several enhancements to the noise bias correction algorithm. We evaluate
    the improved \anacal{} using simulations designed to replicate Rubin LSST
    imaging data. These simulations feature semi-realistic galaxies and stars,
    complete with representative distributions of magnitudes and Galactic
    spatial density. We conduct tests under various observational challenges,
    including cosmic rays, defective CCD columns, bright star saturation, bleed
    trails, and spatially variable point spread functions. Our results indicate
    a multiplicative bias in weak lensing shear recovery of less than a few
    tenths of a percent, meeting LSST DESC requirements without requiring
    calibration from external image simulations. Additionally, our algorithm
    achieves rapid processing, handling one galaxy in less than a millisecond.
\end{abstract}


\section{INTRODUCTION}


We are now entering the era of precision cosmology, marked by the initiation of
groundbreaking astronomical imaging surveys. These surveys will provide precise
tests of the current cosmological paradigm, including understanding the
accelerated expansion rate of the Universe using measurements of cosmic
structure growth with weak gravitational lensing. Gravitational lensing occurs
when the gravity of massive structures in the Universe bends the light from
distant galaxies, resulting in coherent distortions in the images of these
galaxies \citep{rev_wl_Bartelmann01, rev_cosmicShear_Kilbinger15,
rev_wlsys_Mandelbaum2017}. Prominent among these advanced ``stage IV'' imaging
surveys are the Vera C.\ Rubin Observatory Legacy Survey of Space and Time
(LSST, \citealt{LSSTOverviwe2019}), the Euclid mission \citep{Euclid2011}, and
studies using the Nancy Grace Roman Space Telescope \citep{Roman_Spergel2015,
Roman2020}. These imaging surveys require shear measurement techniques that
achieve precision within a few tenths of a percent \citep{WLsys_Massey2013,
LSSTRequirement2018} to meet their ambitious objectives. The deployment of such
sophisticated algorithms necessitates tailored approaches to effectively
process and analyze the data collected by each specific survey.

Several shear measurement techniques have shown promise in meeting the
stringent requirements for analyzing noisy, ``blended'' galaxy images with a
constant applied shear, without the need for calibration through external image
simulations. These methods include \metadet{} \citep{metaDet_Sheldon2020,
metaDet_LSST2023}, a numerical self-calibration method; and \anacal{}
\citep{Anacal_Li2023, Anacal_Li2024}, an analytical framework designed to
derive linear shear response. Of these two, \anacal{} stands out as the most
efficient due to its analytical nature --- it takes $<1$ millisecond to compute
the shear estimator for one detected galaxy. Another promising technique is
\BFD{} \citep{BFD_Bernstein2014, BFD_Bernstein2016}, a Bayesian approach to
shear estimation. However, as of the current writing, it still necessitates
additional percent-level correction, based on simulations, to address detection
biases due to blending. We acknowledge that external image simulations may be
required to calibrate the redshift distribution of source galaxies in cases
where blending occurs between galaxies at different redshifts, each subject to
distinct shear distortions \citep{DESY3_BlendshearCalib_MacCrann2021}. In this
paper, we focus on analytically correcting shear estimation bias under a
constant shear scenario, and defer the discussion of blending effects across
varying redshifts to future work.

\citet{Anacal_Li2023} developed an analytical formalism that effectively
corrects both detection bias and shear estimation bias. This approach, accurate
to second order in shear, operates independent of the specific details of
galaxy morphologies. The method defines the detection process through a series
of step functions based on several basic linear observables. These basic
observables were chosen for their property of having analytically solvable
linear shear responses. By utilizing the shear responses of these basic
observables, \citet{Anacal_Li2023} were able to calculate the linear shear
response of the detection process using the chain rule. The formalism has been
applied to the \FPFS{} shear estimator \citep{FPFS_Li2018, FPFS_Li2022}, a
fixed-kernel method for measuring galaxy shapes, which constructs shapes as
nonlinear functions of shapelet modes. These shapelet modes are obtained by
projecting the galaxy image onto a shapelet basis. We refer readers to
\citet{shapeletsI_Refregier2003, polar_shapelets_Massey2005,
Shapes_Bernstein2002} for details of shapelet basis functions. The shear
response of the galaxy shape is then derived using the chain rule, based on the
pre-calculated shear responses of the shapelet modes.

In \citet{Anacal_Li2023}, a perturbation approach, as outlined in
\citet{KSB_Kaiser2000}, was employed to correct for noise bias in shear
estimation. This method uses the Hessian matrix (which can be fully calculated
using \jax{} \citep{jax_Bradbury2018} following \citealt{Anacal_Li2024}) of the
nonlinear ellipticity, which is defined as a function of basic modes, including
shapelet and peak detection modes. However, this approach necessitates the use
of step functions with very gentle transition for galaxy selection and
detection to maintain stability in the Hessian matrix. Consequently, the
precision of the shear estimation is compromised and remains suboptimal since
it causes ``ambiguous'' detections near the galaxy peaks. In this paper, we
will explore an alternative method for correcting noise bias, as proposed by
\citet{metacal_Huff2017, metacal_Sheldon2017}, which involves adding noise to
galaxy images. We will adapt this approach within the \anacal{} framework,
providing a rigorous analytical proof and introducing several extensions to the
methodology. Although this method internally introduces  additional noise to
the images, it proves to be more effective than the perturbation approach to
noise bias correction. This effectiveness stems from its capability to
incorporate multiple layers of differentiable selection cuts without requiring
the calculation of the noisy Hessian matrix. This approach not only enhances
galaxy detection but also improves the precision of shear estimation.

Furthermore, in this paper, we will demonstrate that the \anacal{} shear
estimator effectively calibrates shear estimates using data that closely
resembles the calibrated LSST coadded images. We generate simulated images
incorporating galaxies and stars, reflecting realistic galactic density
distributions, spatially variable point spread functions (PSF), and proxies for
image artifacts, using the Dark Energy Science Collaboration (DESC)
weak lensing simulation package\footnote{
    \url{https://github.com/LSSTDESC/descwl-shear-sims}
} \citep{metaDet_LSST2023}.
In this paper, our focus is on evaluating the shear estimation technique using
well-calibrated data without errors on the PSF model and the astrometric and
photometric calibrations, and where the  statistics of the image noise are
perfectly known. However, in practical data analyses, calibration inaccuracies
can substantially contribute to systematic errors in shear measurements. It is
important to note that the accuracy of shear measurements may be more dependent
on the quality of data characterization than on the estimation technique
itself.

This paper is organized as follows: Section~\ref{sec:shear_estimator} reviews
the shear estimator and introduces a new analytical noise bias correction
scheme. Section~\ref{sec:sim} presents systematic tests of LSST-like image
simulations under various conditions. Finally, Section~\ref{sec:summary}
provides a summary and outlook.

\section{ANALYTICAL SHEAR ESTIMATION}
\label{sec:shear_estimator}

In this section, we begin by reviewing the derivation of the shear response for
the \FPFS{} shear estimator, which has been calibrated using the \anacal{}
framework (see Section~\ref{sec:method_shear_response}). Following this, we
explore the derivation of the shear response from noisy data, employing two
distinct methodologies (see Section~\ref{sec:method_noise}). The procedure of
our image processing pipeline is detailed in Appendix~\ref{app:pipeline}.

\subsection{Observables and Shear Responses}
\label{sec:method_shear_response}

\begin{figure}
\begin{center}
    \includegraphics[width=0.46\textwidth]{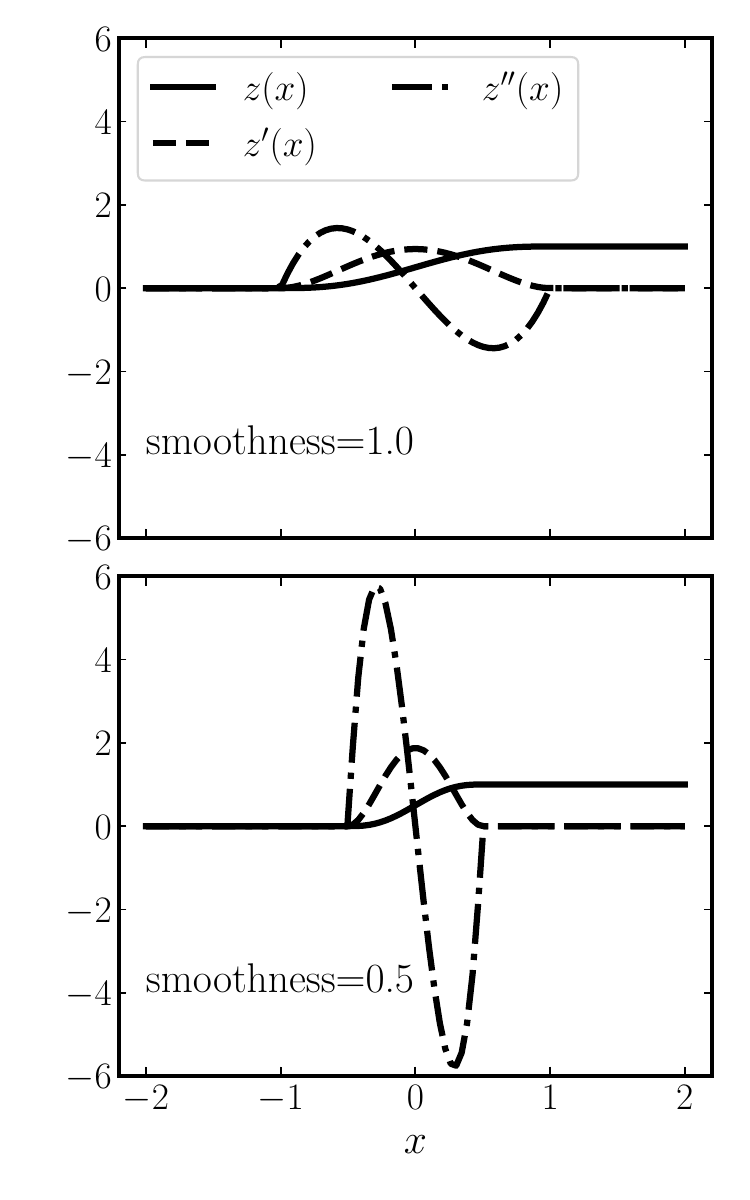}
\end{center}
\caption{
    Smoothstep functions with two smoothness parameters (1 for the upper panel
    and 0.5 for the lower panel) used to select and detect galaxies and their
    derivatives. The step functions ($z(x)$) are shown with solid lines; the
    first ($z'(x)$) and second-order ($z''(x)$) derivatives of the functions
    are in dashed lines and dotted-dashed lines.
    }
    \label{fig:smooth_step}
\end{figure}

We follow \citet{Anacal_Li2023} to use a set of linear observables that are
linear combinations of pixel values to define the galaxy shape, size flux and
detection process. The linear observables include shapelet modes
\citep{shapeletsI_Refregier2003, polar_shapelets_Massey2005} and detection
modes \citep{Anacal_Li2023}. The vector of linear observables are denoted as
$\vnu=(\nu_1, \nu_2, \dots , \nu_n)$, and we follow the notation in
\citet{FPFS_Li2022} --- $\bar{\vnu}$, $\vnu$ and $\tilde{\vnu}$ are the
prelensed, lensed noiseless and lensed noisy observables, respectively.
Additionally, we use a double tilde ($\dtilde{\vnu}$) to denote linear
observables with doubled image noise. We note that these linear observables are
projection of image onto basis kernels after PSF deconvolution in Fourier space
\citep{Z08}. Denoting the basis kernels as $\chi_i(\vk)$, PSF as $p(\vk)$ and
galaxy image as $f(\vk)$, we have
\begin{equation}
    \label{eq:linear_obs_define}
    \nu_i = \int_{\vk} f(\vk) \frac{\chi_i(\vk)}{p(\vk)}\,.
\end{equation}
$\chi_i$ can be any function as long as $\int_{\vk} \frac{\chi_i(\vk)}{p(\vk)}$
is finite. In this paper, we use $\int_{\vk}\!\equiv \int \rmd\vk$ to indicate
integration over the $\vk$ space. In addition, this notation of integration
applies to  any quantities (e.g., $\vnu$).

The basis kernels are composed of polar shapelets and peak detection basis.
Polar shapelets \citep{polar_shapelets_Massey2005}, which are used as basis
functions, are defined in polar coordinates $(\rho, \theta)$ as
\begin{equation}
\label{eq:shapeletfunc}
\begin{split}
\phi_{nm}(\vx \,|\, \sigma_h)&=(-1)^{(n-|m|)/2}\left\lbrace
    \frac{[(n-|m|)/2]!}{[(n+|m|)/2]!}\right\rbrace^\frac{1}{2}\\
    &\times
    \left(\frac{\rho}{\sigma_h}\right)^{|m|}
    L^{|m|}_{\frac{n-|m|}{2}}\left(\frac{\rho^2}{\sigma_h^2}\right)e^{-\rho^2/2\sigma_h^2}
    e^{im\theta},
\end{split}
\end{equation}
where $L^{|m|}_{\frac{n-|m|}{2}}$ are the Laguerre polynomials, $n$ is the
radial number, which can be any non-negative integer, and $m$ is the spin
number, which is an integer between $-n$ and $n$ in steps of two. $\sigma_h$
determines the scale of shapelet functions, which is set to $0\farcs52$ to
maximize the effective galaxy number density according to
\citet{Anacal_Li2024}. Additionally, the peak detection basis functions for the
peak detection modes are
\begin{equation}
\label{eq:peakfuncs_define}
\psi_{i} = \frac{1}{(2\pi)^2} e^{-\abs{\vk}^2\sigma_h^2/2}
\left (
    1 - e^{ \rmi (k_1 x_i  + k_2 y_i) }
\right)
\,,
\end{equation}
where $(x_i, y_i) = (\cos{(i \pi/2)}, \sin{(i \pi/2)})$, and $i=0, 1, 2, 3$\,.
The origin reference point for the measurements of the linear observables is
set at the preselected ``peak candidates'' \citep{Anacal_Li2023}.

To focus on shear estimation, in this paper, we set the lensing convergence to
zero and ignore the distinction between shear and reduced shear. Given that the
lensing shear distortion is small within the weak lensing regime, it is a
practical approach to employ a first-order approximation to understand the
transformation of observables under shear distortion. The linear observables
were chosen due to their advantageous properties, notably that they have
analytically solvable linear shear responses. Specifically, under lensing shear
distortion, the linear observables transform as
\begin{equation}
\label{eq:linear_lensing_shapelets}
\vnu = \left(
    \bar{\vnu}
    + \gamma_1 \vnu_{;1}
    + \gamma_2 \vnu_{;2}
\right)
    + \mathcal{O}(\gamma^2)
    \,,
\end{equation}
where $\vnu_{;1}$ and $\vnu_{;2}$ are the linear shear response with respect
to the first and second components of the lensing shear distortion:
\begin{eqnarray}
    \vnu_{;1} \equiv \frac{\partial \vnu}{\partial \gamma_{1}}\,,
    && \qquad
    \vnu_{;2} \equiv \frac{\partial \vnu}{\partial \gamma_{2}}\,.
\end{eqnarray}
The $i$th components of the linear shear responses with respect to two shear
components are denoted as $\nu_{;1i}$ and $\nu_{;2i}$\,. These observables are
the projection of the deconvolved image onto the shear response of the
corresponding basis functions \citep{Anacal_Li2023}, denoted as
$\chi_{;1i}(\vk)$ and $\chi_{;2i}(\vk)$ for the two shear components:
\begin{eqnarray}
    \nu_{;1i} = \int_{\vk} f(\vk) \frac{\chi_{;1i}(\vk)}{p(\vk)}\,,
    && \quad
    \nu_{;2i} = \int_{\vk} f(\vk) \frac{\chi_{;2i}(\vk)}{p(\vk)}\,.
\end{eqnarray}
More specifically, the shapelet basis set is closed under the shear
perturbation operation. As a result, the shear response of any shapelet basis
can be written as a combination of shapelet basis \citep{Shapes_Bernstein2002}.
The shear responses of the peak detection basis functions are given by
\citet{Anacal_Li2023}:
\begin{equation}
\label{eq:peakfuncs_response}
\begin{split}
\psi_{;1i}
&=
    \frac{1}{(2\pi)^2} e^{-\abs{\vk}^2\sigma_h^2/2}
    (k_1^2-k_2^2)\sigma_h^2
    \left (
        1 - e^{ \rmi (k_1 x_i  + k_2 y_i) }
    \right)\\
    &-
    \frac{1}{(2\pi)^2} e^{-\abs{\vk}^2\sigma_h^2/2}
    \left(
        \rmi x_{i}\,k_1 - \rmi y_{i}\,k_2
    \right)
    e^{ \rmi (k_1 x_i  + k_2 y_i) } \,,\\
    \psi_{;2i}
&=
    \frac{1}{(2\pi)^2} e^{-\abs{\vk}^2\sigma_h^2/2}
    (2\, k_1 k_2)\sigma_h^2
    \left (
        1 - e^{ \rmi (k_1 x_i  + k_2 y_i) }
    \right)\\
    &-
    \frac{1}{(2\pi)^2} e^{-\abs{\vk}^2\sigma_h^2/2}
    \left(
        \rmi y_{i}\,k_1 + \rmi x_{i}\,k_2
     \right)
     e^{ \rmi (k_1 x_i  + k_2 y_i) } \,.
\end{split}
\end{equation}

Nonlinear observables $e_{1,2}$ are defined as nonlinear functions of the
linear observables $\vnu$:
\begin{equation}
    \label{eq:ell_final}
    e_{1,2}(\vnu)
    = \epsilon_{1,2}(\vnu) w_s(\vnu) w_d(\vnu) \,,
\end{equation}
where $\epsilon_{1,2}$ is the spin-2 \FPFS{} ellipticity defined in
\citet{FPFS_Li2018} measuring galaxy shape. $w_s$ represents the selection
weight, while $w_d$ denotes the detection weight \citep{Anacal_Li2023}. They
serve the purposes of galaxy sample selection and galaxy detection,
respectively. These weights are formulated using smoothstep functions of the
linear observables \citep{Anacal_Li2023}. To elaborate, if the weight for a
galaxy is set to zero, it implies rejection from the sample. The smoothstep
functions for a cut at zero are demonstrated in Figure~\ref{fig:smooth_step}.
Since our detection operation, defined with the smoothstep function, is
differentiable, the reference points change smoothly in response to shear
distortion. This allows us to quantify the changes in the population of these
reference points through their shear response. Specifically, under shear
distortion, some reference points may disappear (with their weight
transitioning from non-zero to zero), while others may reappear (with their
weight transitioning from zero to non-zero).

Specifically, the \FPFS{} ellipticity in equation~\eqref{eq:ell_final} is
defined as
\begin{align}
\label{eq:ellipticity_def1}
\epsilon_1 + \mathrm{i}\, \epsilon_2 \equiv \frac{M_{22}}{M_{00}+C}\,,
\end{align}
where the shapelet modes $M_{nm}$ are the linear observables obtained by
projecting (see equation~\eqref{eq:linear_obs_define}) the image onto
deconvolved polar shapelet functions defined in
equation~\eqref{eq:shapeletfunc}. For further details on polar shapelet
functions, we refer readers to \citet{Shapes_Bernstein2002,
polar_shapelets_Massey2005}. The weighting parameter $C$, introduced by
\citet{FPFS_Li2018}, is a constant specific to a galaxy sample. By adjusting
the value of $C$, one can change the relative weights assigned to galaxies of
different brightnesses within the sample. The selection weight in
equation~\eqref{eq:ell_final} has two components --- selection on
signal-to-noise ratio (SNR) denoted as $w_0$ and selection on galaxy size
denoted as $w_2$:
\begin{equation}
    w_s = w_0 \, w_2\,.
\end{equation}
The differentiable selection weight with minimal SNR equals $s_\text{min}$ is
\begin{equation}
\label{eq:sel_cut_snr}
    w_0 =
    z_{\Omega_0} \left( \frac{M_{00}}{\sigma_{0}} - s_\text{min} \right)\,,
\end{equation}
where $\sigma_0$ is the standard deviation of measurement error due to image
noise on the zeroth order shapelet mode $M_{00}$\,. In addition, the
differentiable selection weight with minimal galaxy size $r_\text{min}$ is
\begin{equation} \label{eq:sel_cut_size}
    w_2 =
    z_{\Omega_2} \left( M_{20}+(1 - r_\text{min})M_{00} \right)\,.
\end{equation}
$z_\Omega$ is smoothstep function \citep{smoothstep_Hazimeh20} with smoothness
parameter equals $\Omega$ (see Figure~\ref{fig:smooth_step}):
\begin{equation}
    z_\Omega (x) =
    6 \left(\frac{x + \Omega}{2\Omega} \right)^5
    - 15 \left( \frac{x + \Omega}{2\Omega} \right)^4
    + 10 \left( \frac{x + \Omega}{2\Omega} \right)^3\,,
\end{equation}
where $x\in [-\Omega, \,\Omega]$. The smoothstep function equals zero for
$x<-\Omega$, and 1 for $x>\Omega$\,. In this paper, we adopt two detection
layers. To be more specific, the detection weight is defined as
\begin{equation}
    w_d = z_{\Omega_d}(w_q - w_\text{min})\,,
\end{equation}
where $z_{\Omega_d}$ is the second detection layer, and the first detection
layer is similar to \citet{Anacal_Li2023}:
\begin{equation}
\label{eq:weight_peak1}
    w_q = \prod_{i=0}^{3}\,
    z_{\Omega_q}(q_i - q_\text{min})\,,
\end{equation}
where $q_\text{min} = \Omega_q - 0.8 \sigma_{q}$\,. $q_i$ ($i=0\dots3$) are the
four peak detection modes measuring the difference between neighboring four
pixels after smoothing, and $\sigma_q$ is the standard deviation of measurement
error on the peak detection modes. $q_i$ is obtained by projecting the image
onto deconvolved detection basis functions \citep{Anacal_Li2023}. The second
detection layer performs differentiable smooth selection on the output
detection weight from the first layer to reduce the number of subpeaks near the
galaxy centers. The default hyper parameters are set to $C=4 \sigma_0$\,,
$\Omega_0\!=\!\Omega_2\!=\!1.6 \sigma_0$\,, $\Omega_q\!=\!1.6\sigma_q$\,,
$\Omega_d\!=\!0.04$, $s_\text{min}\!=\!12$, $r_\text{min}\!=\!0.1$,
$w_\text{min}\!=\!0.12$\,. These hyperparameters are fine-tuned to optimize the
effective galaxy number density using image simulations. We have confirmed that
the accuracy remains robust even when the parameters are varied slightly around
their optimal values. We refer readers to \citet{Anacal_Li2024} for more
details on how to optimize the effective galaxy number density.

Before presenting the linear shear response of the weighted ellipticity
$e_{1,2}$, we conduct a spin analysis to justify that the first order
approximation is accurate to the second order of shear due to the rotational
symmetry. Initially, we note that the galaxy ellipticity $\epsilon_{1,2}$ is
inherently a spin-2 quantity so that it negates under a $90\degr$ rotation. The
selection and detection weights are specifically designed to remain invariant
under a $90\degr$ rotation, as outlined by \citet{Anacal_Li2023}. As a result,
the weighted ellipticity reverses sign under a $90\degr$ rotation indicating
the absence of both spin-0 and spin-4 components. Assuming that the intrinsic
galaxies are isotropically oriented and their positions are randomly
distributed prior to shear distortion, since the weighted ellipticity lacks
spin-0 components, we have
\begin{equation}
\label{eq:zero_deriv_wrt_shear}
    \left\langle \left. e_{1,2} \right|_{\vec{\gamma} = 0} \right\rangle_g = 0,
\end{equation}
where $\langle \bigcdot \rangle_g$ is the average over galaxy population.
Additionally, since the weighted ellipticity lacks both spin-0 and spin-4
components, we have
\begin{equation}
\label{eq:secondorder_deriv_wrt_shear}
    \left\langle \left.
    \frac{\partial^2 e_{1, 2}}{\partial\gamma_i \partial\gamma_j}
    \right |_{\vec{\gamma} = 0}
    \right\rangle_g
    = 0,
\end{equation}
where $i, j \in \{1, 2\}$ (see Appendix~A of \citealt{Anacal_Li2023}).
Furthermore, from equation~\eqref{eq:secondorder_deriv_wrt_shear}, we derive
\begin{equation}
\left\langle
\frac{\partial  e_i}{\partial \gamma_j}
\right\rangle =
\left\langle \left.
    \frac{\partial  e_i}{\partial \gamma_j}
    \right |_{\vec{\gamma} = 0} \right\rangle_g
    + \mathcal{O}(\gamma^2)\,,
\end{equation}
indicating that we can directly assess the shear response to the first order in
shear from the distorted image without reverting to the intrinsic image before
shear distortion. The bias introduced by this estimation of the shear response
is at the second-order of shear. Therefore, the expectation value of the
weighted ellipticity under the lensing shear distortion can be derived with a
Taylor expansion and omitting the zeroth order and the second-order terms since
their expectation values are zero:
\begin{equation}
\label{eq:shear_transform}
    \langle e_i \rangle = \sum_{j=1}^{2}
    \left\langle\frac{\partial  e_i}{\partial \gamma_j}\right\rangle_g \gamma_j
    + \mathcal{O}(\gamma^3)\,,
\end{equation}
where $i=1,2$\,. We observe that equation~\eqref{eq:shear_transform} remains
accurate up to the second order of shear, as the expectation value of the
weighted ellipticity's second-order derivatives with respect to shear is zero.
For more details, we refer readers to \citet{Anacal_Li2023}.

The linear shear response matrix, $\frac{\partial  e_i}{\partial \gamma_j}$,
where $j=1,2$, can be expressed with the linear shear response of the linear
observables using the chain rule:
\begin{equation}
\label{eq:chain_rule}
    \frac{\partial  e_i}{\partial \gamma_j} = \sum_k
    \frac{\partial  e_i}{\partial \nu_k}
    \frac{\partial  \nu_k}{\partial \gamma_j}
    = \sum_{k} \frac{\partial  e_i}{\partial \nu_k} \nu_{;j k}\,,
\end{equation}
where $\nu_{;j k}$ is the $k$th component of the shear response vector
$\vnu_{;j}$\,.

In the following subsections, we will derive the shear response for the first
component of ellipticity for noisy data. The methodology for the second
component follows the same approach and can be derived accordingly.

\subsection{Estimating Shear Response From Noisy Data}
\label{sec:method_noise}

In real observations, there is noise in the images, causing measurement error.
We denote the measurement error on the linear observable vector as $\delta
\vnu$, and we have
\begin{equation}
    \delta\nu_i = \int_{\vk} n(\vk) \frac{\chi_i(\vk)}{p(\vk)}\,,
\end{equation}
where $n(\vk)$ is the pure noise image in Fourier space, and the integration is
over the 2D Fourier space. The nonlinear function $e_1$ is calculated on the
noisy linear observables $\tilde{e}_1= e_1\!(\tilde{\vnu})$, where
$\tilde{\vnu} = \vnu + \delta \vnu$. The noise is described by its probability
density function (PDF) $P_n(\delta \vnu)$. We focus on the condition that the
sky background is precisely subtracted such that the sum of the noise PDF
equals one and it possesses a zero mean:
\begin{eqnarray}
    \int_{\delta \vnu} P_n(\delta \vnu) = 1,
    && \qquad \int_{\delta \vnu} \delta \nu_i P_n(\delta \vnu) = 0\,.
\end{eqnarray}
where $\nu_i$ is the $i$th element of the linear observable vector, and the
integration on the noise realization over the space defined with the states of
the image noise. Furthermore, in the derivation, we assume that the image noise
is homogeneous and additive, which is a reasonable approximation for faint,
small galaxies observed in ground-based surveys because the main source of
noise comes from fluctuations in the sky background. However, it is important
to note that this assumption does not hold in practical scenarios when
interpolating over bad pixels, cosmic rays, etc. As discussed in
Section~\ref{sec:sim_artifact}, we address this by masking pixels with
artifacts prior to detection and shear estimation.

The average ellipticity measured from noisy data is
\begin{equation}
\label{eq:ell_noisy_1}
\left\langle \tilde{e}_1 \right\rangle
    = \int_{\vnu} P_g(\vnu) \int_{\delta \vnu} e_1\!(\vnu + \delta\vnu)
      P_n(\delta \vnu) \,,
\end{equation}
where $P_g(\vnu)$ is the galaxy PDF after the lensing effect. Due to the
invariance of the probability measure --- the probability of the random
variable falling within a particular volume before and after the transformation
should be the same --- we have $P_g(\vnu) \rmd \vnu = \bar{P}_g(\bar{\vnu})
\rmd \bar{\vnu}$, where $\bar{P}_g$ is the PDF before lensing shear distortion.
In this paper, PDFs with bar refer to those before the lensing shear
distortion. Additionally, $\langle \bigcdot \rangle$ is the average over galaxy
population and noise realizations.

We note that due to the nonlinearity in ellipticity, the average galaxy
ellipticity over different noise realizations is biased compared to the
noiseless estimation:
\begin{equation}
\int_{\delta \vnu} e_1\!(\vnu + \delta\vnu) P_n(\delta\vnu)
\neq e_1\!(\vnu)\,,
\end{equation}
and the corresponding bias in shear estimation is termed noise bias
\citep{noiseBiasRefregier2012}.
Without any noise bias correction, our shear estimation exhibits a
multiplicative bias of a few percent, which is approximately ten times higher
than the subpercent-level requirement set by the LSST DESC.

In the following context of this subsection, we outline two distinct analytical
approaches for noise bias correction. These methodologies do not make any
assumption on galaxy morphology and do not rely on simulated images. The
first approach involves correcting the noise bias in the ensemble average of
the noisy ellipticity, transforming it back to that expected for the noiseless
ellipticity, which is then used for shear measurement
(Section~\ref{sec:method_noise_correct}). The second approach directly
estimates the shear response based on the expectation value of the noisy
ellipticity (Section~\ref{sec:method_noise_renoise}).

\subsubsection{Noise Correction Approaches}
\label{sec:method_noise_correct}

This methodology draws its inspiration from the foundational work of
\citet{KSB_Kaiser2000}, and has been refined in \citet{FPFS_Li2022} and
\citet{Anacal_Li2023}. Specifically, we compute the Taylor expansion of the
noisy ellipticity as a function of $\delta \vnu$, offering a derivation
of shear bias to 2nd order in image noise:
\begin{equation}
\label{eq:ell_taylor_expand}
\begin{split}
e_1\!(\vnu + \delta\vnu)
    & = e_1\!(\vnu)
    + \frac{\partial e_1\!(\vnu)}{\partial \nu_i} \delta\nu_i
    + \frac{1}{2}
    \frac{\partial^2 e_1\!(\vnu)}{\partial \nu_i
    \partial \nu_j} \delta\nu_i \delta\nu_j \\
    & + \mathcal{O}(\delta \nu^3)\,.
\end{split}
\end{equation}
We direct readers to equation~(21) of \citet{FPFS_Li2022} for a detailed
application of this noise bias correction method. Here we employ Einstein
notation, wherein repeated indices imply summation over those indices. We
substitute equation~\eqref{eq:ell_taylor_expand} into
equation~\eqref{eq:ell_noisy_1} and the expectation value of the noisy
ellipticity is
\begin{equation}
\label{eq:noise_correct_2nd_e}
\begin{split}
\left\langle \tilde{e}_1 \right\rangle
    & = \langle e_1 \rangle
    + \frac{1}{2}
    \left\langle
        \frac{\partial^2 e_1\!(\vnu)}{\partial \nu_i \partial \nu_j}
    \right\rangle_g
    \left\langle \delta \nu_i \delta \nu_j \right\rangle_n
    + \mathcal{O}(\delta \nu^3) \,,
\end{split}
\end{equation}
where $\langle \bigcdot \rangle_g$ is the average over lensed noiseless galaxy
population (operation $\int_\vnu P_g(\vnu) \lbrack \bigcdot \rbrack$); whereas
$\langle \bigcdot \rangle_\text{n}$ is the average over noise realizations
(operation $\int_{\delta \vnu} P_n(\delta \vnu) \lbrack \bigcdot \rbrack$).
$\left\langle \delta \nu_i \delta \nu_j \right\rangle_n$ is the covariance
matrix of measurement error on the linear observables that are given in
\citet{Anacal_Li2023}. We note that the above derivation assumes the
realization of the image noise (the value of the image noise) is independent of
the specifics of the galaxy’s surface brightness profile. This assumption holds
true for ground-based surveys that are dominated by sky background. However, it
is not valid in space-based observations which are dominated by source Poisson
noise, where the variance of the image noise is dependent on the expectation of
the pixel value. Another source of image noise is readout noise from the
detector. When the readout noise is homogeneous across the measurement scale
(as defined by the shapelet kernel), this correction scheme can be applied to
mitigate noise bias. While additional calibration may be required to address
any inhomogeneity in the noise distribution, such considerations are beyond the
scope of this paper.

The corrected expectation value of ellipticity $\langle e \rangle$ is distorted
by the lensing shear, and it is related to the PDF of the galaxy population
before shearing:
\begin{equation}
    \langle e_1 \rangle =
    \int_{\vnu} P_g(\vnu) e_1\!(\vnu) =
    \int_{\bar{\vnu}} \bar{P}_g(\bar{\vnu}) e_1\!\left(T(\bar{\vnu})\right)\,,
\end{equation}
where $\bar{P}_g$ is the PDF of galaxy population before the shear distortion,
and $T$ is the mapping from the intrinsic linear observables $\bar{\vnu}$ to
the lensed linear observables $\vnu$. Following equation~\eqref{eq:chain_rule},
the linear shear response at the single galaxy level can be written as a
function of linear observables
\begin{equation}
R_1(\vnu)
=
    \frac{\partial e_1\!(\vnu)}{\partial \nu_i}
    \frac{\partial{\nu_i}}{\partial \gamma_1}
=
    \frac{\partial e_1\!(\vnu)}{\partial \nu_i}
    \nu_{;1i}
    \,.
\end{equation}
Here, we present only the diagonal term of the linear shear response matrix, as
the expectation value of the off-diagonal terms is zero \citep{Anacal_Li2023}.
When dealing with noisy data, we first measure the expectation of the noisy
shear response $\left\langle \tilde{R}_1\right\rangle$, and then apply a
second-order noise bias correction to obtain $\left\langle R_1\right\rangle$
(the expectation of the second term on the right-hand side of
equation~\eqref{eq:ell_taylor_expand}):
\begin{equation}
    \left\langle R_1 \right\rangle
=
    \left\langle \tilde{R}_1 \right\rangle
    - \frac{1}{2}\left\langle
    \frac{\partial^2 R(\vnu)}{\partial \nu_i \partial \nu_j}
    \right\rangle
    \left\langle \delta \nu_i \delta \nu_j \right\rangle_n
    + \mathcal{O}(\delta \nu^3)
    \,.
\end{equation}
The shear estimator is
\begin{equation}
\label{eq:shear_estimator_perturb}
    \hat{\gamma}_1 = \frac{\langle e_1 \rangle}
    {\left\langle R_1 \right\rangle}
    + \mathcal{O}(\gamma^3) + \mathcal{O}(\delta \nu^3)\,.
\end{equation}
We observe that this method necessitates the calculation of second- and higher-
order derivatives of the step function. The function's sharpness is modulated
by the smoothness parameter; a larger parameter results in a gentler
transition. However, a caveat is that smaller smoothness parameters, which
create a sharper selection function, lead to significant fluctuations in these
higher order derivatives (see the lower panel of Figure~\ref{fig:smooth_step}).
As a result, we are constrained to employing only smoothly transitioned cuts
for our source detection and galaxy selection, which may lead to ``ambiguous''
detections near the peak of galaxy sources \citep{Anacal_Li2023}.

\subsubsection{Renoising Approach}
\label{sec:method_noise_renoise}

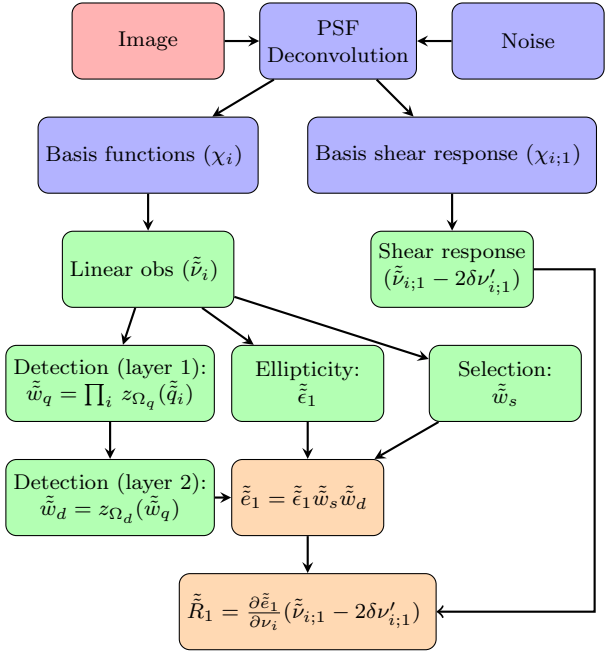
\begin{figure}
\nolinenumbers
\centering
\begin{tikzpicture}[node distance=1.5cm]
    \node (image) [data] {Image};
    \node (psf) [filter, right of = image, xshift=1cm] {
        PSF\\
        Deconvolution
    };
    \node (noise) [filter, right of = psf, xshift=1cm] {Noise};
    \node (filter) [filter, below of= image] {
        Basis functions ($\chi_i$)
    };
    \node (filter response) [filter, right of= filter, xshift=2.5cm] {
        Basis shear response ($\chi_{i;1}$)
    };
    \node (linObs) [obs, below of= filter] {
        Linear obs ($\dtilde{\nu}_i$)
    };
    \node (linRes) [obs, right of= linObs, xshift=2.5cm] {
        Shear response \\
        ($\dtilde{\nu}_{i;1} - 2\delta{\nu}'_{i;1}$)
    };
    \node (wdet1) [obs, below of= linObs, xshift=-0.5cm] {
        Detection (layer 1): \\
        $\dtilde{w}_{q}=\prod_i \, z_{\Omega_q}(\dtilde{q}_i)$
    };
    \node (wdet2) [obs, below of= wdet1] {
        Detection (layer 2):\\
        $\dtilde{w}_{d}=z_{\Omega_d}(\dtilde{w}_{q})$
    };
    \node (ell) [obs, right of= wdet1, xshift=1.1cm] {
        Ellipticity:\\
        $\dtilde{\epsilon}_1$
    };
    \node (wsel) [obs, right of= ell, xshift=1.1cm] {
        Selection:\\
        $\dtilde{w}_s$
    };
    \node (well) [res, right of= wdet2, xshift=1.1cm] {
        $\dtilde{e}_1 = \dtilde{\epsilon}_1 \dtilde{w}_s \dtilde{w}_d$
    };
    \node (jac) [res, below of= well] {
        $\dtilde{R}_1 =
        \frac{\partial \dtilde{e}_1}{\partial \nu_i}
        (\dtilde{\nu}_{i;1} - 2\delta{\nu}'_{i;1})$
    };

    \draw [arrow] (image) -- (psf);
    \draw [arrow] (noise) -- (psf);
    \draw [arrow] (psf) -- (filter);
    \draw [arrow] (psf) -- (filter response);
    \draw [arrow] (filter) -- (linObs);
    \draw [arrow] (filter response) -- (linRes);
    \draw [arrow] (linObs) -- (wdet1);
    \draw [arrow] (linObs) -- (ell);
    \draw [arrow] (linObs) -- (wsel);
    \draw [arrow] (wdet1) -- (wdet2);
    \draw [arrow] (wdet2) -- (well);
    \draw [arrow] (wsel) -- (well);
    \draw [arrow] (ell) -- (well);
    \draw [arrow] (well) -- (jac);
    \draw [->, thick] (linRes.east) -- ++(0.8,0) -- ++(0.0,-4.5) -- (jac.east);
\end{tikzpicture}
    \caption{
        The workflow of \anacal{}-\FPFS{} from the input image to the output
        ellipticity and its shear response. To simplify the workflow, we only
        demonstrate the estimator for the first component of the shear. The
        workflow of our pipeline is summarized in Appendix~\ref{app:pipeline}.
    }
    \label{fig:workflow}
\end{figure}
\begin{figure}
\begin{center}
    \includegraphics[width=0.46\textwidth]{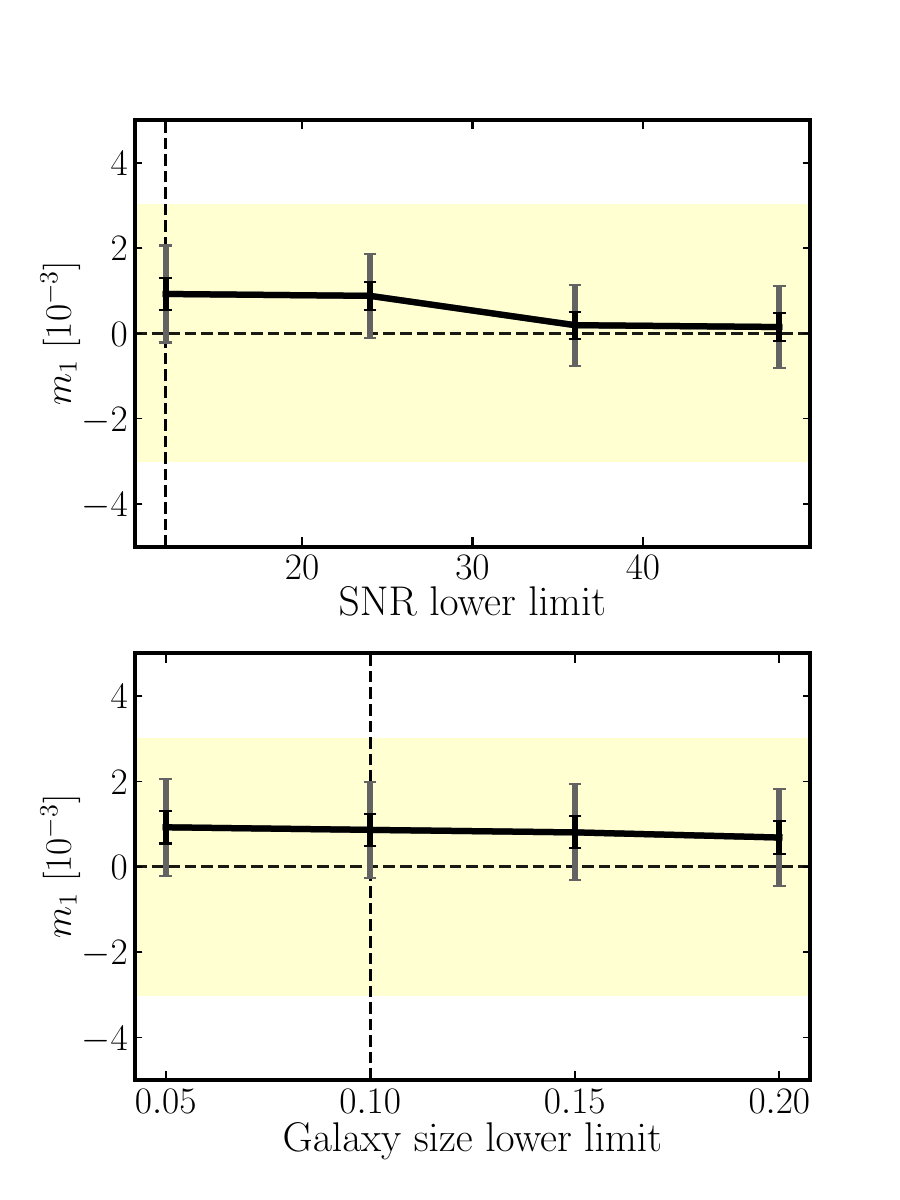}
\end{center}
\caption{
    Multiplicative bias with 1$\sigma$ and 3$\sigma$ errors as a function of
    SNR lower limit (upper panel) and resolution lower limit (lower panel). The
    shaded region shows the LSST ten-year requirement on the control of
    multiplicative bias \citep{LSSTRequirement2018}. The vertical lines are the
    default selection cuts on SNR (SNR$>12$) and galaxy size ($R_2>0.1$).
    }
    \label{fig:test_basic}
\end{figure}

The technique of incorporating noise into images to develop a shear estimator
free from noise bias, as outlined in this subsection, was first introduced by
\citet{metacal_Sheldon2017}. The advantage of this method is that it does not
require second- and higher-order derivatives of the nonlinear observables for
noise bias correction. Furthermore, there is no need to approximate by ignoring
the bias from higher-order noise. However, \citet{metacal_Sheldon2017} did not
provide a formal proof for the algorithm, and how it differs from the
perturbation approach undertaken in \citet{Anacal_Li2023}. In this study, we
present an analytical formalism for correcting noise bias using renoised data,
which builds on the foundational framework established by
\citet{Anacal_Li2023}. This advancement allows us to enhance our understanding
of the noise bias correction technique proposed by \citet{metacal_Sheldon2017},
resulting in a significant increase in processing speed compared to the
original numerical approach documented in \citet{metacal_Sheldon2017}.

Following \citet{metacal_Sheldon2017}, we add an additional layer of noise to
the image, which possesses identical statistical properties after being rotated
counterclockwise by $90\degr$, where the rotation is specified within the space
before PSF smearing. The resulting additional measurement error on the linear
observables are denoted as $\delta \vnu'$, and they are characterized by a PDF
denoted as $P_n^{90}(\delta \vnu')$. This $90\degr$ rotation is applied to
eliminate any spin-2 anisotropies present in the original noise image after
deconvolution, which may stem from either the anisotropy in the noise
correlation function or the PSF anisotropy leakage in the process of PSF
deconvolution. Consequently, the ellipticity observed post-application of the
additional noise image is:
\begin{equation}
\label{eq:ell_double_noisy_1}
\begin{split}
    \left\langle \dtilde{e}_1 \right\rangle
=&
    \int_{\bar{\vnu}, \delta \vnu, \delta \vnu'}
    \bar{P}_g(\bar{\vnu})
    \,
    e_1\!\!\left(T(\bar{\vnu}) + \delta\vnu + \delta\vnu' \right) \\
    &\times
    P_n(\delta \vnu) P_n^{90}(\delta \vnu')
    \,.
\end{split}
\end{equation}

To simplify the analysis, we define a new nonlinear observable:
\begin{equation}
\begin{split}
    G_1 (T(\bar{\vnu})) &=
    \int_{\delta\vnu, \delta\vnu'}
    e_1\!\!\left(T(\bar{\vnu}) + \delta \vnu + \delta \vnu'\right)
    \\
    & \times
    P_n\!\left(\delta\vnu\right)
    P_n^{90}\!\left(\delta\vnu'\right)\,,
\end{split}
\end{equation}
and equation~\eqref{eq:ell_double_noisy_1} can be rewritten as
\begin{equation}
\left\langle \dtilde{e}_1 \right\rangle
    = \int_{\vnu} \bar{P}_g(\bar{\vnu}) G_1(T(\bar{\vnu})) \,.
\end{equation}
Again, before presenting the shear response of the renoised ellipticity
$\dtilde{e}_{1,2}$, we conduct a spin analysis to justify that the first-order
approximation is accurate to the second order of shear. As we assume that the
noise is homogeneous, its statistical properties remain the same under a
$180\degr$ rotation with respect to any reference point. Consequently, $P_n
P_n^{90}$ is invariant under a $90\degr$ rotation since the statistical
property of the doubled noise remain unchanged under a $90\degr$ rotation.
Additionally, given that $e_1$ negates under a $90\degr$ rotation (see
Section~\ref{sec:method_shear_response}), it follows that $G_1(\nu)$ also
negates under a $90\degr$ rotation. Therefore, $G_1(\vnu)$ lacks both spin-0
and spin-4 components, necessitating solely the computation of its linear shear
response to achieve a shear estimator accurate up to the second order of shear
(see equations~\eqref{eq:zero_deriv_wrt_shear} and
\eqref{eq:secondorder_deriv_wrt_shear}).

To derive the shear response of the renoised ellipticity, we first determine
the shear response of $G_1(T(\bar{\vnu}))$ (adopting Einstein notation):
\begin{equation}
\label{eq:shear_response_G}
\begin{split}
    \left.
    \frac{\partial G_1(T(\bar{\vnu}))}{\partial \gamma_1}
    \right|_{\vec{\gamma}=0}
=
    &
    \int_{\delta \vnu, \delta \vnu'}
    \frac{\partial e_1\!(\dtilde{\vnu})}{\partial \dtilde{\nu}_i}
    \,
    \frac{\partial{\dtilde{\nu}_{i}}}{\partial \gamma_1}
    \\
    & \left.\times P_n\!\left(\delta{\vnu}\right)
    P_n^{90}\!\left(\delta{\vnu'}\right)
    \right|_{\vec{\gamma} =0}
    \,,
\end{split}
\end{equation}
where $\dtilde{\vnu}$ is the linear observable measured from the renoised image
--- $\dtilde{\vnu} = T(\bar{\vnu})+ \delta\vnu + \delta\vnu'$\,. Since image
noise is not affected by lensing shear, the shear responses of $\delta \vnu$
and $\delta \vnu'$ are zero. Consequently, we have:
\begin{equation}
\partial{\dtilde{\nu}_{i}}/ \partial \gamma_1 = \partial{\nu_{i}}/ \partial
\gamma_1 =\nu_{;1i}\,.
\end{equation}
However, $\nu_{;1i}$ cannot be directly obtained in real observations due to
the presence of noise in the images. The observed, renoised observable is:
\begin{equation}
    \dtilde{\nu}_{;1i} = \int_{\vk} \dtilde{f}(\vk)
    \frac{\chi_{;1i}(\vk)}{p(\vk)}\,,
\end{equation}
where $\dtilde{f}(\vk)$ is the renoised image in Fourier space. Similarly, we
have the corresponding quantity defined for the image noise:
\begin{equation}
    \delta \nu_{;1i} = \int_{\vk} n(\vk)
    \frac{\chi_{;1i}(\vk)}{p(\vk)}\,.
\end{equation}
Thus, we can derive:
\begin{equation}
    \frac{\partial \dtilde{\nu}_i}{\partial \gamma_1} = \nu_{;1i}
    = \dtilde{\nu}_{;1i} - \delta{\nu}_{;1i} - \delta{\nu}'_{;1i}\,,
\end{equation}
and the shear response of the expectation of the renoised ellipticity is
\begin{equation}
\left\langle \dtilde{R}_1 \right\rangle
=
    \left\langle
    \left.
    \frac{\partial e_1\!(\dtilde{\vnu})}{\partial \dtilde{\nu}_i}
    \,
    \left(
    \dtilde{\nu}_{;1i} - \delta \nu_{;1i} - \delta \nu'_{;1i}
    \right)
    \right|_{\vec{\gamma} = 0}
    \right\rangle\,.
\end{equation}
Note that $\delta \vnu'$ can be measured because the noise is generated through
simulation and we know the exact noise realization; in contrast, $\delta \vnu$,
which represents the noise in actual images, cannot be directly measured since
reconstructing the exact noise realization from the image is unfeasible.
However, given that $\delta \vnu'$ and $\delta \vnu$ are from noise
realizations oriented $90 \degr$ relative to each other in the pre-PSF plane,
and considering the random orientation of galaxies, it follows that:
\begin{equation}
\label{eq:noise_response}
    \left\langle
    \left.
    \frac{\partial e_1\!(\dtilde{\vnu})}{\partial \dtilde{\nu}_i}
    \right|_{\vec{\gamma} = 0}
    \delta \nu'_{;1i}
    \right\rangle
=
    \left\langle
    \left.
    \frac{\partial e_1\!(\dtilde{\vnu})}{\partial \dtilde{\nu}_i}
    \right|_{\vec{\gamma} = 0}
    \delta \nu_{;1i}
    \right\rangle\,.
\end{equation}
We refer readers to Appendix~\ref{app:noise_bias} for the proof. Therefore, the
expectation of the shear response of the renoised ellipticity can be measured
with the simulated noise:
\begin{equation}
\label{eq:shear_response_final}
\left\langle \dtilde{R}_1 \right\rangle
=
    \left\langle
    \frac{\partial e_1\!(\dtilde{\vnu})}{\partial \dtilde{\nu}_i}
    \,
    \left(
    \dtilde{\nu}_{;1i} - 2 \delta \nu'_{;1i}
    \right)
    \right\rangle\,,
\end{equation}
and the shear estimator is
\begin{equation}
\label{eq:shear_estimator_renoise}
    \hat{\gamma}_1 = \frac{\langle \dtilde{e}_1 \rangle}
    {\left\langle \dtilde{R}_1 \right\rangle} + \mathcal{O}(\gamma^3)\,.
\end{equation}
In summary, by introducing the additional noise, we eliminate the spin-2
anisotropy from the original image noise on the image plane prior to PSF
convolution. Furthermore, we utilize this added noise, whose realization are
exactly known, to estimate the shear response of the renoised image. It is
worth noting that, the ``renoising'' approach eliminates the need for any
truncation in the perturbative series for the noise bias correction.
Consequently, equation~\eqref{eq:shear_estimator_renoise} does not include the
high-order terms from image noise, $\mathcal{O}(\delta \nu^3)$, which are
present in equation~\eqref{eq:shear_estimator_perturb}.

Both the perturbation and ``renoising'' approaches involve trade-offs between
variance and bias. For the perturbation method, the second-order noise bias
correction term for ellipticity, measured from a noisy image, increases the
statistical uncertainty in the shear estimation. Similarly, in the ``renoising''
approach, the additional noise doubles the variance of the image noise, also
increasing the variance in shear estimation. As a result, the effective galaxy
number density is reduced by 20\% to 25\% compared to using a biased shear
estimator without noise bias correction.

The workflow of the algorithm from input image to the output shape and the
corresponding shear response is demonstrated in Figure~\ref{fig:workflow} and
outlined in Appendix~\ref{app:pipeline}. We note that this noise bias
correction does not involve second- or higher-order derivatives during the
detection process. This allows us to apply a second layer of detection using
step functions with small smoothness parameters to avoid ``ambiguous''
detection (see the layer 2 in Figure~\ref{fig:workflow}). To be more specific,
in the first detection layer, we measure the preliminary detection weight:
$\dtilde{w}_q$ using the value difference between neighboring pixels to select
peak candidates following \citet{Anacal_Li2024}. The second layer applies
another step function on $\dtilde{w}_p$ to derive the final detection weight:
$\dtilde{w}_d = z_{\Omega_d}(\dtilde{w}_q)$, where $z_{\Omega_d}$ is the step
function with a small smoothness parameter $\Omega_d = 0.04$\,. We find that,
due to this improved detection, the effective galaxy number density is
improved by approximately $15\%$ compared to the earlier version of the code
adopting the second-order noise bias correction in \citet{Anacal_Li2024}. More
specifically, we have observed that the second-order noise bias correction term
for ellipticity in equation~\eqref{eq:noise_correct_2nd_e} becomes particularly
noisy when either the selection weight or the detection weight has a large
gradient. As illustrated in Figure~\ref{fig:smooth_step}, the noise bias
correction terms that incorporate second-order derivatives display significant
fluctuations under conditions of selection weight function with a small
smoothness parameter. This instability leads to considerable uncertainty in
shear estimation in noisy images. Moreover, employing a detection weight
function with a large smoothness parameter results in numerous sub-peaks near
the galaxy center where the weight is nonzero, which cannot be applied to real
observations. Additionally, it is worth noting that the ``renoising'' approach
does not require any noisy second-order derivatives so that we are able to
adopt multiple detection layers with small smoothness parameters. Moreover, it
avoids the need for any truncation in the perturbative series for the noise
bias correction. Therefore, we adopt the ``renoising'' approach as the default
noise bias correction method in \anacal{}.

However, a limitation of this noise bias correction method is that we need to
double the image noise before detection and measurement. It worth noting that a
recent paper \citep{deepfield_metacal2023} suggests using deep field images
from the same survey to mitigate the image noise in such correction strategies.
We intend to apply the methodology of \citet{deepfield_metacal2023} in our
analytical shear estimation framework in future work.

\section{TEST ON IMAGE SIMULATIONS}
\label{sec:sim}

\begin{figure}
\begin{center}
    \includegraphics[width=0.46\textwidth]{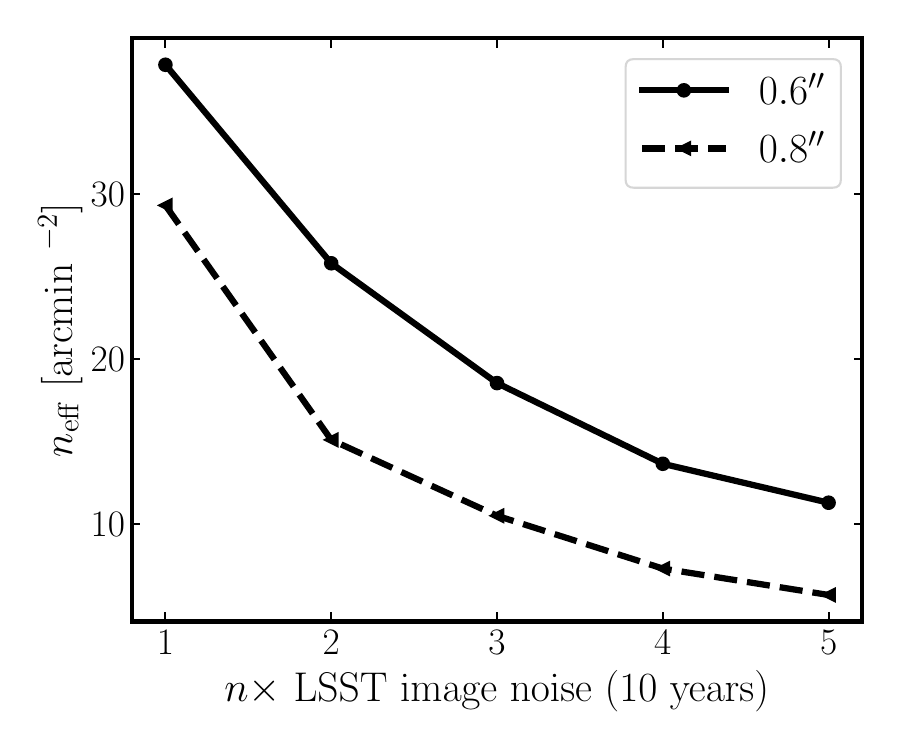}
\end{center}
\caption{
    This figure illustrates the effective galaxy number density, normalized so
    that the effective standard deviation of shape noise per component is 0.26.
    The data combines information from four bands: ``$griz$''. The calculation
    of the effective galaxy number density is based on the anticipated noise
    level of the ten-year LSST coadds. The solid line (dashed) is for the
    PSF seeing 0\farcs6 (0\farcs8). Actual observations may yield different
    values.
    }
    \label{fig:test_neff}
\end{figure}

In this section, we evaluate our shear estimation algorithm using LSST-like
image simulations under various conditions, as detailed in
Section~\ref{sec:sim_sim}. The results for different galaxy sample selections
are presented in Section~\ref{sec:sim_gal}. Section~\ref{sec:sim_psf_e}
examines the effects of different PSF anisotropies, while
Section~\ref{sec:sim_psf_var} discusses the impact of varying levels of PSF
variation across the coadded image cells. The influence of anisotropy in the
noise correlation function is detailed in Section~\ref{sec:sim_noise_e}.
Section~\ref{sec:sim_star} addresses the results under various levels of
stellar contamination. Finally, Section~\ref{sec:sim_artifact} evaluates the
algorithm’s performance in the presence of bright stars and pixel artifacts.

\subsection{Simulation}
\label{sec:sim_sim}

\begin{figure}
\centering
\includegraphics[width=0.46\textwidth]{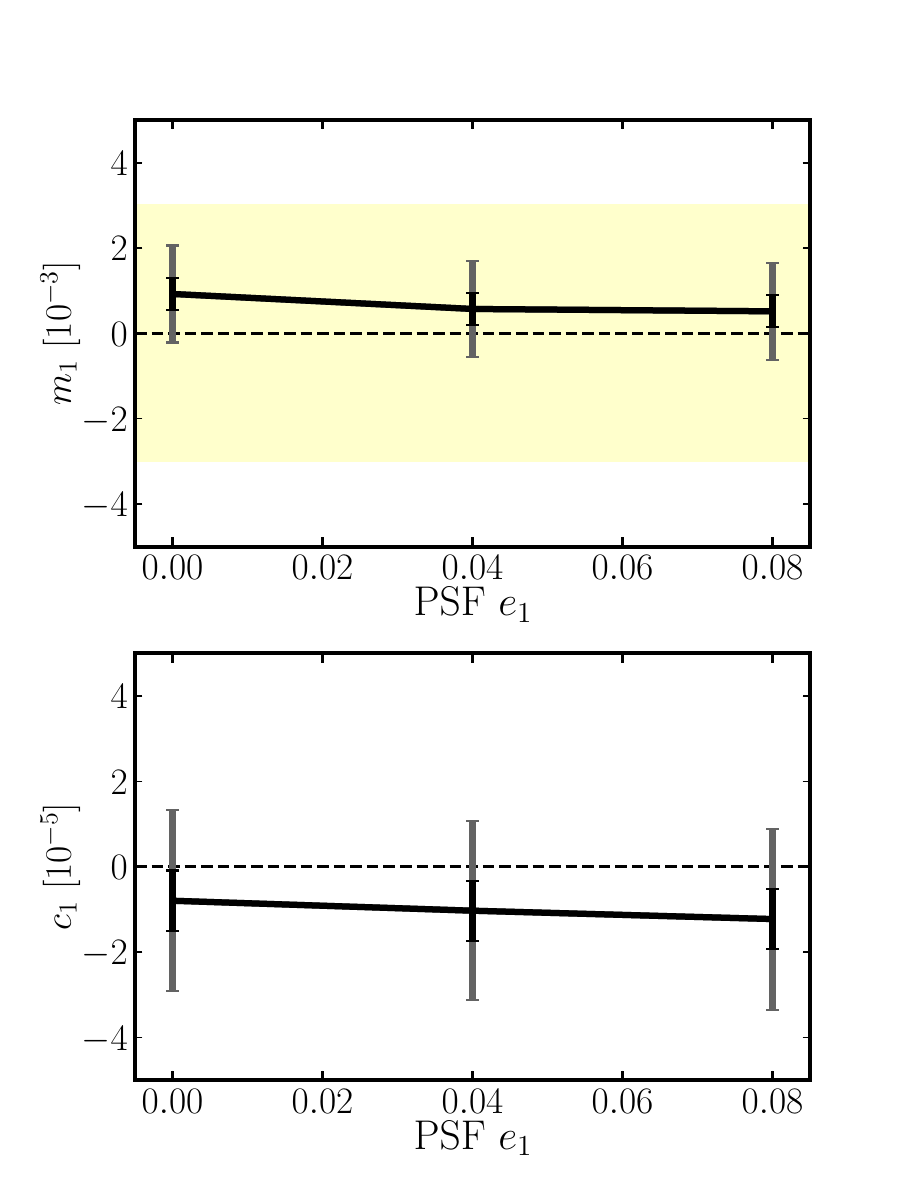}
\caption{
    Multiplicative bias (upper panel) and additive bias (lower panel) with
    1$\sigma$ and 3$\sigma$ errors as a function of PSF anisotropy. The shaded
    region are the LSST ten-year requirement on the control of systematic bias
    \citep{LSSTRequirement2018}.
    }
    \label{fig:test_psf_e}
\end{figure}

\begin{figure}
\centering
\includegraphics[width=0.46\textwidth]{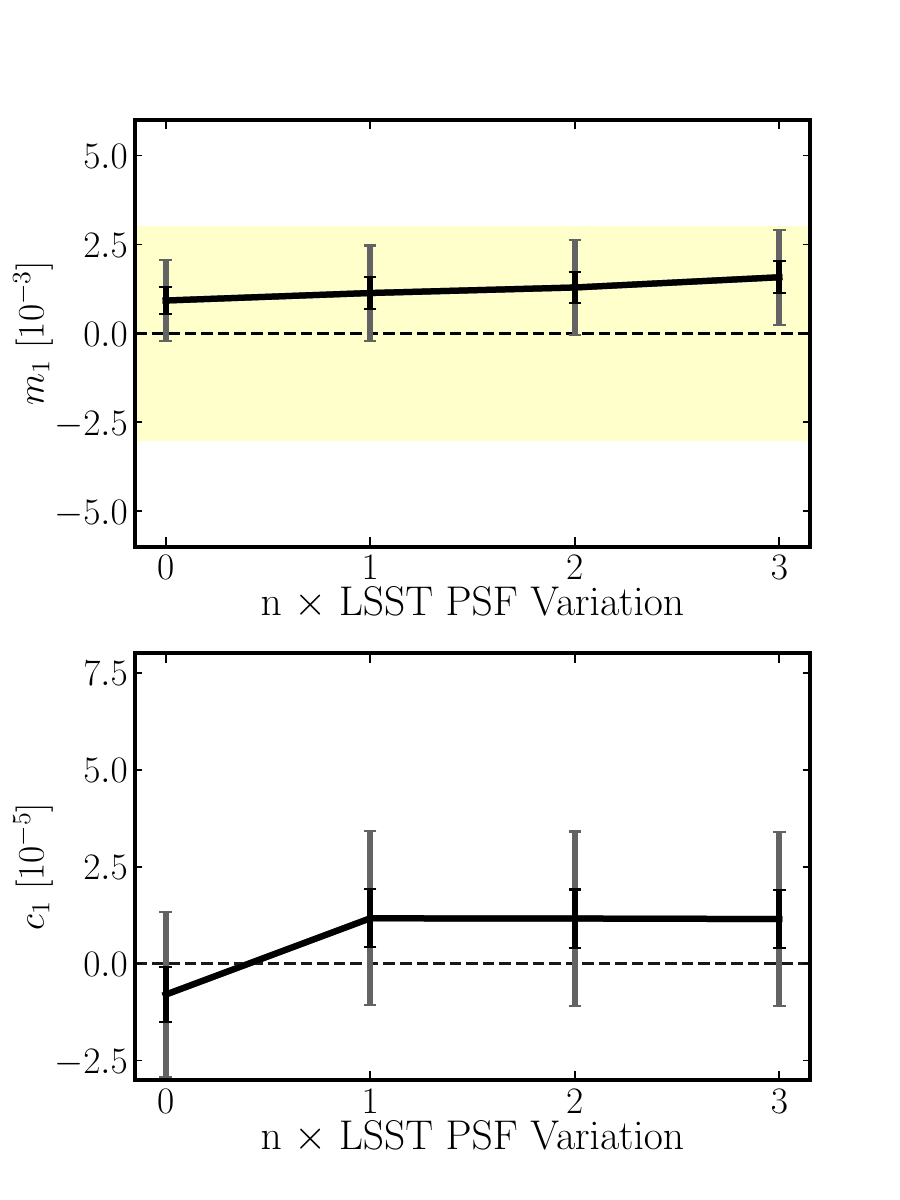}
\caption{
    Multiplicative bias (upper panel) and additive bias (lower panel) as a
    function of PSF variation, with 1$\sigma$ and 3$\sigma$ errors. The shaded
    region shows the LSST ten-year requirement on the control of multiplicative
    bias \citep{LSSTRequirement2018}.
    }
    \label{fig:test_psf_var}
\end{figure}

We employ the package \texttt{descwl-shear-sims}\footnote{
    \url{https://github.com/LSSTDESC/descwl-shear-sims}
} as presented in \citet{metaDet_LSST2023} for image simulations. This package
renders images using the \galsim{} package \citep{GalSim}. The simulation
package generates calibrated images with subtracted background and provides an
estimated noise variance image plane. Additionally, it includes a world
coordinate system (WCS) transformation and a position-dependent point spread
function (PSF) model. Areas of concern such as saturation, star bleeds, cosmic
rays, and defective columns are identified and flagged within an integer bit
mask image plane for easy reference. We do not simulate scenarios involving
miscalibrated input data. Specifically, we omit testing the effects of
inaccuracies in PSF models, noise estimates, image warping, image coadding,
astrometric or photometric calibrations. Our primary aim is to evaluate the
effectiveness of our shear estimator using coadd images whose characteristics
are well understood. However, to fully assess the final shear calibration in
real data analyses, it might become essential to either propagate or simulate
the impact of miscalibrated data. We note that all of the simulations that are
used to test our shear estimation includes galaxy blending. Additionally the
simulations adopt a constant shear, where all blended galaxies within a single
image experience identical shear distortions. Both the galaxy shapes and the
positions are changed by the shear. The shear applied in these simulations is
$\gamma_1 = \pm 0.02$\,. The simulation is divided into $10000$ subfields for
each test case, and each subfield is 0.06 square degrees. We adopt the ring
test \citep{galsim_STEP2} by creating a $90\degr$ rotated companion for each
subfield. Additionally, we use the shape noise cancellation proposed by
\citet{preciseSim_Pujol2019} to reduce uncertainty on the estimated shear cased
by intrinsic shape dispersion and image noise.

Our basic image simulation adopts an LSST-like PSF, for which the PSF image is
modelled with a \cite{Moffat1969} profile:
\begin{equation}
\label{eq:moffat_PSF}
    p_{m}(\vx)=\left[1+c\left(\frac{|\vx|}{r_\mathrm{P}}\right)^2\right]^{-2.5}\,,
\end{equation}
where $r_\mathrm{P}$ and $c$ are set such that the full width at half maximum
(FWHM) of the Moffat PSF is $0\farcs80$, matching the expected median seeing of
LSST images. Note that we start with a circular PSF and examine asymmetric PSFs
later in Section~\ref{sec:sim_psf_e}.

Due to the high sky background, we simulate the residual Poisson noise in each
image by applying Gaussian noise after the background subtraction. We utilized
the \texttt{WeakLensingDeblending}\footnote{
    \url{https://github.com/LSSTDESC/WeakLensingDeblending}
} package \citep{weaklensDeblend} to estimate the image noise tailored
to the LSST filter specifications. After rendering all image features and
introducing noise, we standardized the images to a consistent zero point of 30.

In this paper, we simulate coadded images in the ``$griz$'' bands using the
anticipated ten-year LSST noise levels as our standard simulation setup, unless
otherwise noted. The input galaxy models are generated using the
\texttt{WeakLensingDeblending} package. These models feature galaxies with
bulge, disk, and AGN components, all sharing the same morphology across
different bands. The bulge and disk components can have different fluxes and
ellipticities, and the AGN is modeled as point source at galaxy center. The
catalog boasts a raw density of 240 galaxies per square arcminute and has an
effective i-band AB magnitude limit of 27. Galaxies are randomly distributed in
the image without taking into account clustering of galaxy positions.

We quantify the bias in the shear estimation using multiplicative ($m_{1,2}$)
and additive ($c_{1,2}$) biases \citep{shearSys_Huterer2006, Heymans2006}.
Specifically, the estimated shear, $\hat{g}_{1,2}$, is related to the true
shear, $g_{1,2}$, as
\begin{equation}
\label{eq:shear_biases}
    \hat{g}_{1,2}=(1+m_{1,2})\,g_{1,2}+c_{1,2}\,.
\end{equation}
We adopt $90$ degree rotation \citep{galsim_STEP2} and the technique introduced
in \citet{preciseSim_Pujol2019} to reduce shape noise in the estimation of
multiplicative and additive biases.

\subsection{Galaxy Properties}
\label{sec:sim_gal}

\begin{figure}
\centering
\includegraphics[width=0.35\textwidth]{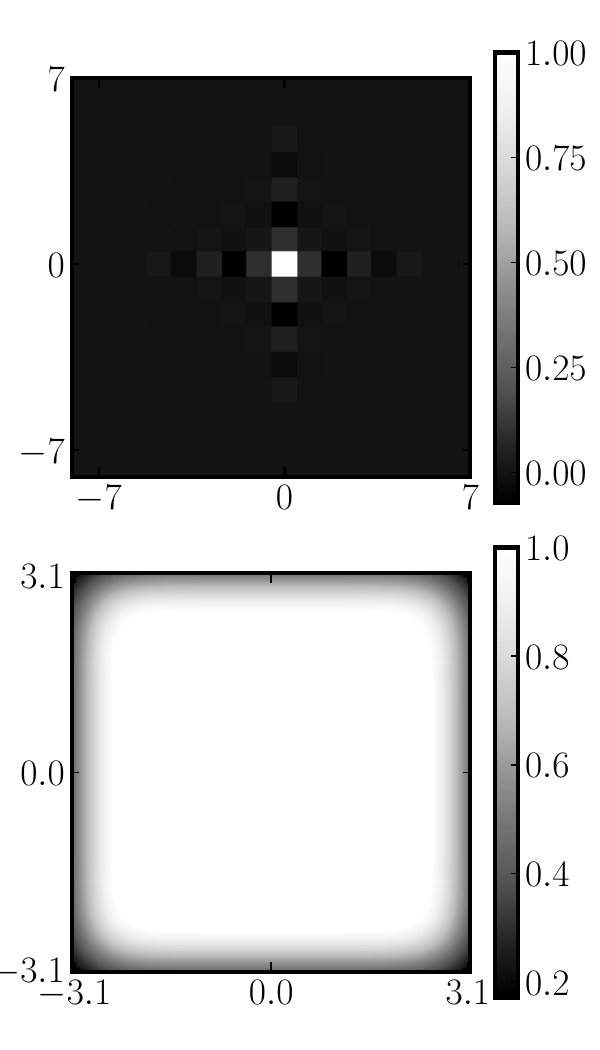}
\caption{
    The two-point statistics of homogeneous Gaussian noise. The upper panel
    shows the noise correlation function in configuration space, and the lower
    panel shows the power spectrum in Fourier space within the Nyquist
    wavenumber. These two-point statistics are normalized such that the maximum
    is $1$. For illustration purposes in the figure, the pixel scale is set to
    1, resulting in a Nyquist wavenumber of $\pi$\,.
    }
    \label{fig:noise_2pt}
\end{figure}

In this subsection, we demonstrate the performance of our shear estimator using
simple image simulations characterized by isotropic PSFs and uniform noise
levels. The PSF remains constant across the image, and we avoid complications
from image artifacts such as bad CCD columns or cosmic rays. Moreover, these
simulations only include galaxies and omit stars. We focus on testing the
performance of the shear estimator with different selection cuts on galaxy SNR
and size.

In Figure~\ref{fig:test_basic}, we present the multiplicative bias as a
function of the lower limit of galaxy SNR in the upper
panel, and as a function of the lower limit of galaxy size in the lower panel.
We find that the multiplicative biases for these cuts are all below the LSST
science requirement on the control of multiplicative shear bias
\citep{LSSTRequirement2018}. Since no significant additive bias is detected in
this analysis, it is not included in the plots. We refer readers to
\citet{Anacal_Li2023} for the definition of galaxy size ($R_2$) and SNR. In the
remainder of this paper, unless otherwise specified, we adopt SNR$>\!\!12$,
$R_2\!>\!0.1$ as the default criteria for galaxy selection.

Additionally, we have estimated the effective galaxy number density for varying
noise levels. Our definition of effective galaxy number density is normalized
such that the effective standard deviation of shape noise per component is
$0.26$. Our estimation strictly adheres to equation~(5) from
\citet{Anacal_Li2024}. Specifically, we simulate numerous exposures with
blended galaxies (with different galaxy populations, orientations and image
noises) and obtain the shear estimation uncertainty using all detections within
one square arcminute. The uncertainty in shear estimation is denoted as
$\sigma_g$. We then calculate the number of galaxies with an intrinsic shape
noise of 0.26 required to achieve this level of uncertainty, determining the
estimated effective galaxy density per square arcminute.
The effective galaxy number density is
\begin{equation}
    \label{eq:effective_number}
    n_\text{eff} = \left(\frac{0.26}{ \sigma_g} \right)^2
    ~[\mathrm{arcmin}^{-2}]\,.
\end{equation}
Compared to other methods of measuring effective galaxy number density by
counting galaxies with specific weights \citep{WLsurvey_neffective_Chang2013},
our approach accounts for the correlation between shape measurements of
neighboring detections. This correlation arises because a single image pixel
can contribute to the measurements of multiple neighboring galaxies when they
are blended. The corresponding results are presented in
Figure~\ref{fig:test_neff}. This analysis utilizes data from the four-band
``$griz$'' combination, and the ``renoising'' approach is adopted for noise
bias correction. In the simulations, the galaxy morphology and the PSF model
are exactly the same across all bands. Similar to \citet{Anacal_Li2024}, we
combine images from different bands by adding them with inverse-variance
weights. It is important to note that the effective galaxy number density is
estimated based on the expected noise levels from ten years of LSST coadds and
the anticipated median PSF size. However, the absolute value of the effective
number density may not be realistic since we have not precisely calibrated the
number density in the simulations to any real dataset.

\subsection{Anisotropy from PSF}
\label{sec:sim_psf_e}

In this subsection, we show the performance of our shear estimator on image
simulations with different values for the PSF anisotropy. To simulate these
images, the Moffat PSF model defined in equation~\eqref{eq:moffat_PSF} is
sheared to exhibit an ellipticity of $(e_1=e_{\mathrm{psf}}, e_2 =
-e_{\mathrm{psf}})$\,, where $e_{\mathrm{psf}}$ represents the PSF anisotropy,
varying between $0$ and $0.08$\,. For the PSFs with small ellipticities,
the relation between ellipticity and shear is approximately: $e_{1,2}\sim 2
g_{1,2}$\,.

The results depicted in Figure~\ref{fig:test_psf_e} demonstrate that changes in
PSF anisotropy do not significantly impact either the multiplicative or
additive shear biases. The multiplicative bias remains below the LSST requirement,
and the additive bias is statistically consistent with zero. Notably, even in
the most extreme scenario with $e_{\mathrm{psf}}=0.08$, the magnitude of the
additive bias remains below $3 \times 10^{-5}$. We also calculate the
fractional additive bias, defined as the ratio of the additive bias to the PSF
ellipticity \citep{HSC1_GREAT3Sim, HSC3_catalog}:
\begin{equation}
    a_{1,2} \equiv \frac{c_{1,2}}{e^{\mathrm{PSF}}_{1,2}}\,,
\end{equation}
and we report a fractional additive bias of $a_1 = (1.6 \pm 1.1) \times
10^{-4}$, which is consistent with zero.

\subsection{PSF Variation}
\label{sec:sim_psf_var}

\begin{figure}
\centering
\includegraphics[width=0.46\textwidth]{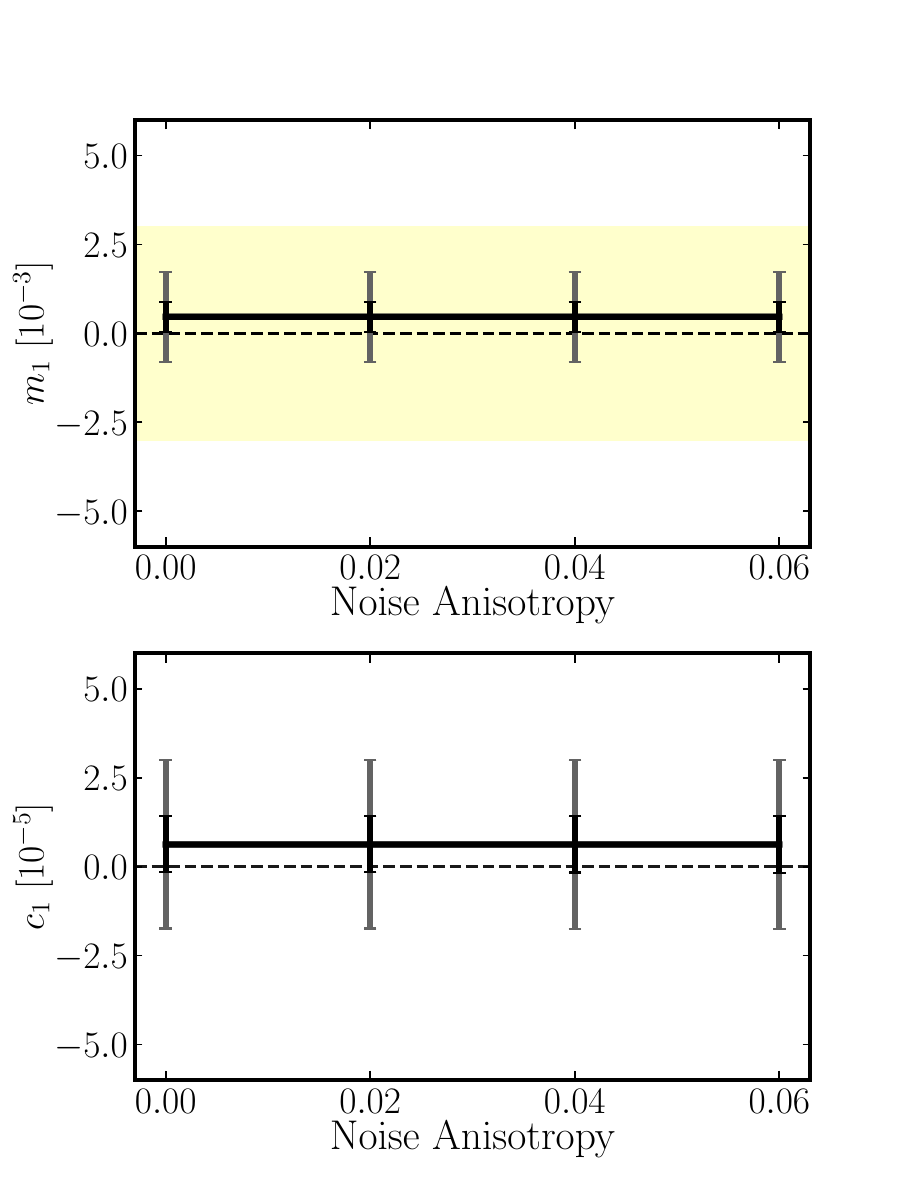}
\caption{
    Multiplicative bias (upper panel) and additive bias (lower panel) with
    1$\sigma$ and 3$\sigma$ errors as a function of noise anisotropy
    (quantified by the shear distortion on noise correlation function). The
    shaded region shows the LSST ten-year requirement on the control of
    multiplicative bias \citep{LSSTRequirement2018}.
    }
    \label{fig:test_noise_e}
\end{figure}

\begin{figure}
\centering
\includegraphics[width=0.46\textwidth]{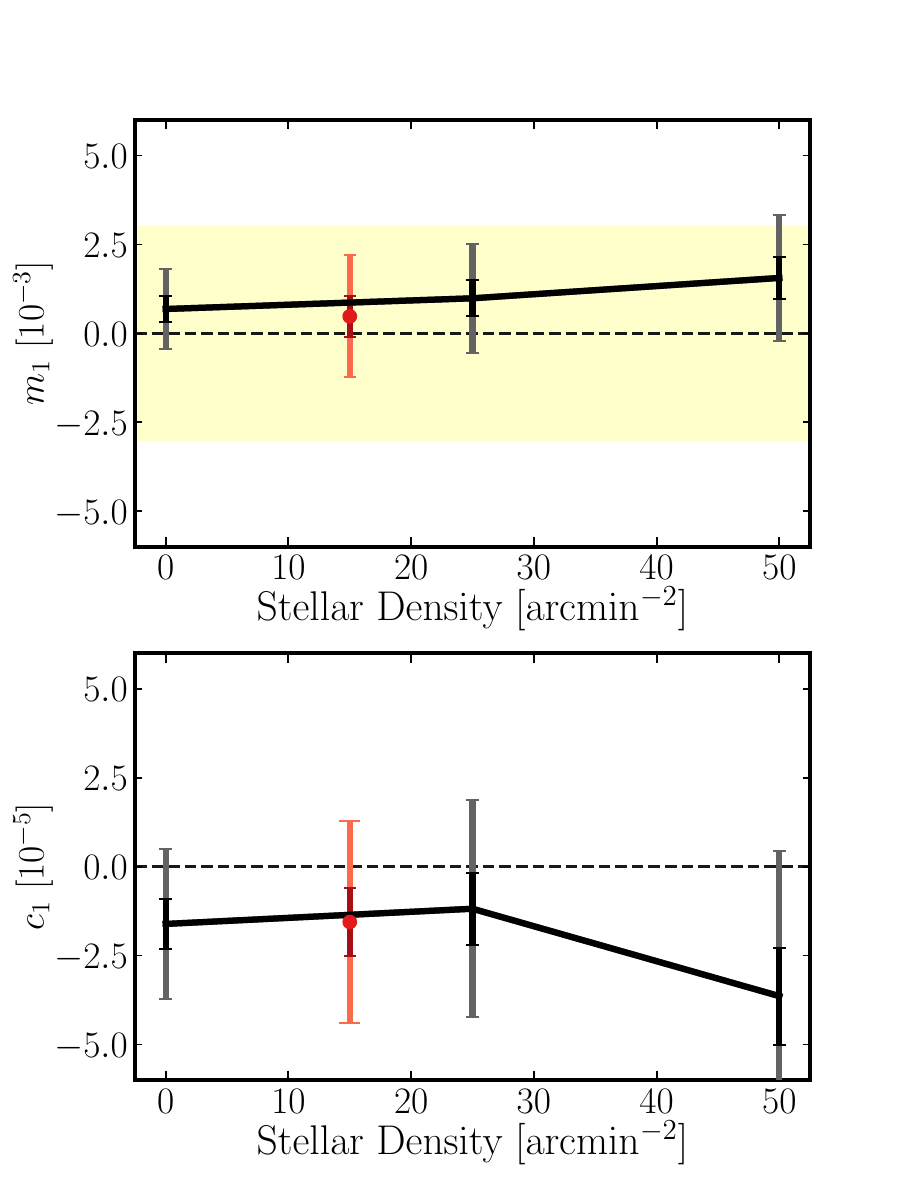}
\caption{
    Multiplicative bias (upper panel) and additive bias (lower panel) with
    1$\sigma$ and 3$\sigma$ errors as a function of stellar density in the
    simulation. The ``$\bullet$'' points indicate the biases, as established
    through simulations where the stellar density for each subfield is randomly
    drawn from the DC2 simulation across the LSST survey. The shaded region
    shows the LSST ten-year requirement on the control of multiplicative bias
    \citep{LSSTRequirement2018}.
    }
    \label{fig:test_stellar_density}
\end{figure}

We create spatially variable PSF models by modifying the basic Moffat PSF model
(equation~\eqref{eq:moffat_PSF}), varying both the ellipticity and size of the
PSFs across the image. The simulation methods are detailed in Appendix A of
\citet{metaDet_Sheldon2020} and build upon the approach proposed by
\citet{PSFVar_Heymans2012}. \citet{PSFVar_Heymans2012} utilized images with
high stellar density to fit a von Kármán model of atmospheric turbulence
\citep{PSFVar_vonK1948} to account for PSF variations. Meanwhile,
\citet{metaDet_Sheldon2020} refined this algorithm by reducing unrealistically
high power below one arcsecond, thus generating realizations of spatially
variable PSFs using Gaussian random fields. In the image simulation, we
model the PSF ellipticity and size at different positions as a Gaussian random
field, using correlation functions from \citet{PSFVar_Heymans2012}. To account
for variations, we use a PSF variation parameter that modulates the
variance of the Gaussian random field to simulate PSF field with different
levels of variation.

In our detection and shape measurement process, we employ a constant point
spread function (PSF) across each coadded image cell
\citep{LSST_coaddCell2024}, which measures 250 by 250 pixels. This constant PSF
is sampled at the center of the image cell. Specifically, our shear inference
approach does not capture spatial variations of the PSF below scales of
$\sim$250 pixels. Each neighboring cell includes an overlapping region of 50
pixels. Figure~\ref{fig:test_psf_var} illustrates the performance of the shear
estimator for different levels of PSF variation. In
Figure~\ref{fig:test_psf_var}, we normalize the variation of the PSF by the
expected PSF variance for cell-based coadds in the final year of the Rubin LSST
survey \citep{metaDet_Sheldon2020}. We observe that the multiplicative bias
increases slightly with greater PSF variation, whereas the additive bias
remains consistently zero. It is important to note that the multiplicative bias
arises from using a constant PSF sampled at the center of each cell. Even at a
variance three times that expected in the LSST, the multiplicative bias remains
within acceptable LSST standards. The results for estimation of the second
component of the shear is consistent with the first component of shear
demonstrated in Figure~\ref{fig:test_basic}.

\subsection{Anisotropic Noise}
\label{sec:sim_noise_e}

In this subsection, we evaluate our algorithm's performance in the presence of
noise with various anisotropic correlations due to the image warping. Operating
under the assumption that the noise correlation function can be precisely
measured in real observations, our aim is to determine whether we can mitigate
the bias introduced by correlated noise, provided that we have an accurate
assessment of its correlation function. Testing the performances of image
warping and coadding are beyond the scope of this paper.

We focus on anisotropic (square-like) correlations between adjacent pixels,
which match the autocorrelation function of a third-order ($a=3$) Lanczos
kernel, which was used to warp and coadd images taken during the first-year HSC
survey
\citep{HSC1_pipeline}:
\begin{equation}
    L(\vec{x})\!=\!\begin{cases}
        \sinc{\frac{x}{a}}\,\sinc{x}\,\sinc{\frac{y}{a}}\,\sinc{y}
        & \! \abs{x},\abs{y}<a\\
        0
        & \! \mathrm{otherwise,}
    \end{cases}
\end{equation}
where $\sinc{x}=\sin{(\pi x)}/\pi x$; $x$, $y$ and $a$ are in pixel units.
Since the third-order Lanczos kernel is undersampled for the pixel size in the
observation, we estimate the correlation function using a finer pixel scale
where the Lanczos kernel is oversampled, which is equivalent to estimating the
correlation function from the coadded noise images from single exposure with
random subpixel offsets after warping with the Lanczos kernel. The correlation
function and power spectrum below the Nyquist wavenumber are shown in
Figure~\ref{fig:noise_2pt}. The noise correlation function exhibits a pattern
similar to the average noise correlation function derived from the blank pixels
in the HSC coadded images, as illustrated in Figure 1 of
\citet{HSC1_shapecalib}. This correlation function shows a stronger variation
along the $x$ and $y$ axes, which can be attributed to the properties of the
Lanczos kernel. Unlike some observations (e.g., HSC and LSST), where single
exposures undergo different rotations, no rotation is applied when generating
the effective noise correlation function. The power spectrum is computed using
a Discrete Fourier Transform of the correlation function. To prevent the
influence of periodic boundary conditions in the computation of Discrete
Fourier Transform, the stamp size for the Fourier transform is set
significantly larger than the size of the Lanczos kernel. In addition, the
Fourier transform is performed on the finer pixels to reduce the alising effect
on the power spectrum due to the undersampling. In the simulation, we generate
a normalized white noise field and filter it according to the noise power
spectrum below the Nyquist wavenumber. We then rescale the noise field to match
a target noise variance. The noise applied in this test is homogeneous and
Gaussian since (a) inhomogeneous noise originating from galaxy sources plays a
minor role and (b) the high sky background level in ground-based observations,
such as Rubin LSST, means the Poisson noise from the background is effectively
Gaussian.

\begin{figure*}
\centering
\includegraphics[width=0.9\textwidth]{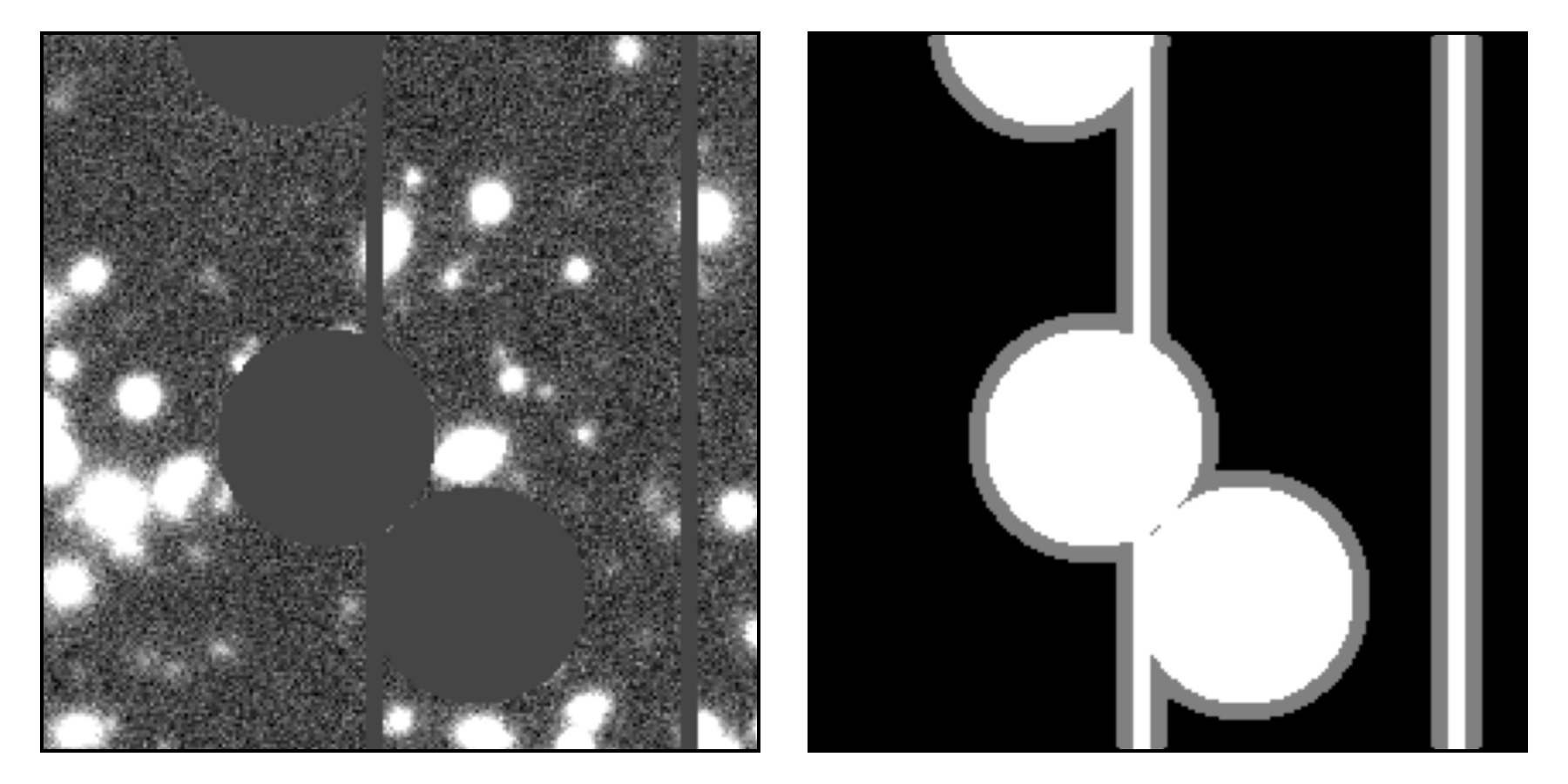}
\caption{
    The left panel displays the image plane with a mask applied prior to the
    detection and measurement process. The right panel illustrates the mask
    plane itself. In the right panel, the white pixels represent regions that
    were masked before detection and measurement. The grey pixels indicate
    areas within the extended mask where sources have a mask value exceeding 20
    (see vertical lines in Figure~\ref{fig:test_mask}). These sources are
    subsequently excluded from the catalog.
    }
    \label{fig:test_msk_img_demo}
\end{figure*}

To simulate various noise anisotropy levels, we distort the noise correlation
function using a shear ranging from $0$ to $0.06$ to stretch the noise
correlation function in the horizontal direction. In the image processing, we
assume that we know the exact noise correlation function after the distortion.
In Figure~\ref{fig:test_noise_e}, we depict both multiplicative and additive
biases as functions of the shear distortion applied to the noise correlation
function. As demonstrated, the biases in the shear estimator remain
consistently consistent with zero.

\subsection{Stellar Contamination}
\label{sec:sim_star}

This subsection assesses the impact of stellar contamination on shear
estimation. As demonstrated in \citet{FPFS_Li2022}, when
sources are isolated, both the ellipticity and shear response expectation
values for a star sample are zero. Thus, while stellar contamination does
introduce variance to the estimates, it does not bias them in cases involving
isolated sources. However, this observation cannot be extended to situations
where sources are blended. In blended scenarios, the position of
stars---predominantly from the Milky Way---is unaffected by lensing
distortions. Consequently, when stars and galaxies are blended, it leads to
contamination of the shear signal in distant galaxies.

To evaluate the accuracy of shear estimation in scenarios characterized by
blending with stellar density distributions, we performed simulations using
blended images. These simulations incorporated stars with fluxes and densities
akin to those found in the Milky Way, as derived from the LSST DESC Data
Challenge Two (DC2) simulation catalogs, as reported by \citet{DescDC2_2021}.
In each simulated field, stars are randomly sampled with replacement from the
stellar density map utilized in the DC2 simulation, with densities exceeding
100 per square arcminute being excluded. For this test, we omitted stars
brighter than magnitude 18 in $r$-band, while in the subsequent section, we
will specifically examine the impact of the bright-star mask on shear
estimation. It is important to note that the specified stellar density refers
to the total number of stars input into the simulation, not those actually
detected. Each star is modeled as a point source convolved with a PSF.

Results from this test are indicated by ``$\bullet$'' points in
Figure~\ref{fig:test_stellar_density}, showing that both multiplicative and
additive shear biases are consistently close to zero. The average stellar
density for these samples is approximately 18 arcmin$^{-2}$. Further detailed
tests are conducted with fixed input stellar densities ranging from 0 to 50
arcmin$^{-2}$. These results, also shown in
Figure~\ref{fig:test_stellar_density}, indicate that even at an extremely high
stellar density (50 arcmin$^{-2}$), neither the multiplicative nor the additive
biases are significant (i.e., less than $3\sigma$). Moreover, the
multiplicative shear bias remains below the LSST's requirements, even in this
extreme case.

\subsection{Bright Star and Pixel Masks}
\label{sec:sim_artifact}

In this subsection, we evaluate the performance of the shear estimator in
simulations that include bright stars and image artifacts such as saturation,
bleeds, bad CCD columns, and cosmic rays. The simulation follows
\citet{metaDet_LSST2023}.

We continue to randomly sample stars from the stellar density map used in the
DC2 simulation. Unlike Section~\ref{sec:sim_star}, which only includes stars
fainter than magnitude 18, this simulation includes all stars, even those that
are brighter. To mimic saturation, we limited the value in each pixel, flagged
saturated pixels in the integer bitmask image of the exposure, and set the
variance for these pixels to infinity. Although we did not simulate nonlinear
detector responses, we overlaid saturated stars with bleed trail images from
pre-generated templates that match the star’s flux in the relevant filter.
We flag these pixels with bleed trails appropriately in the bitmask.

For cosmic rays, we randomly determined their location, angle, and length
(between 10 and 30 pixels) on the image. We flagged the affected pixels and
ensured adjacent pixels were also flagged if they touched corners. These pixels
were set to NaN in the image data, and a cosmic ray bit is added to the
bitmask for later interpolation, preventing their use in the final shear
estimates.

Additionally, we used a modified Monte Carlo generator from Becker et. al (in
prep.) to simulate bad columns. Each column, just a pixel wide and randomly
positioned, included random gaps to reflect columns that do not extend across
the entire CCD. Each image included at least one such bad column. The bad
columns are flagged in the bitmask.

Our masking algorithm is divided into two steps. In the first step, we apply
circular binary masks to stars with a magnitude brighter than 18, extending out
to the radius where the star’s profile met the $1\sigma$ noise level of the
image. It is important to note that in real-world scenarios, such masks would
be algorithmically determined. These masked regions near bright stars are set
to zero. Pixels affected by image artifacts, including star bleeds, cosmic
rays, and bad columns, are also zeroed out. The left panel of
Figure~\ref{fig:test_msk_img_demo} shows a masked image before detection and
measurement, and the right panel shows the mask plane.

In the second step, we aim to identify sources in close proximity to the edge
of the mask that have been significantly impacted by the masking process
setting the pixels in the masked region to zero. Specifically, we make a
continuous smoothed version of this binary mask in the first step and record
the mask value from the smoothed mask at the position of the detected peak. A
higher mask value indicates closer distance to the masked region and greater
influence of the masking on the source. Then we eliminate detections at pixels
falling above a threshold in this smoothed mask. The smoothing kernel is a 2D
Gaussian. with a standard deviation matching the scale of the shapelet kernel
used in the \FPFS{} shear estimator \citep{FPFS_Li2018}. The Gaussian kernel is
truncated at three times its standard deviation. The Gaussian kernel is
normalized to a flux of 1000, and we retain only the integer part of the mask
value to save storage space. Essentially, we increase the mask region and
remove the detection in the extended masked region. We note that the mask value
is not dependent on lensing shear distortion; therefore, this selection does
not cause selection bias.

\begin{figure}
\centering
\includegraphics[width=0.46\textwidth]{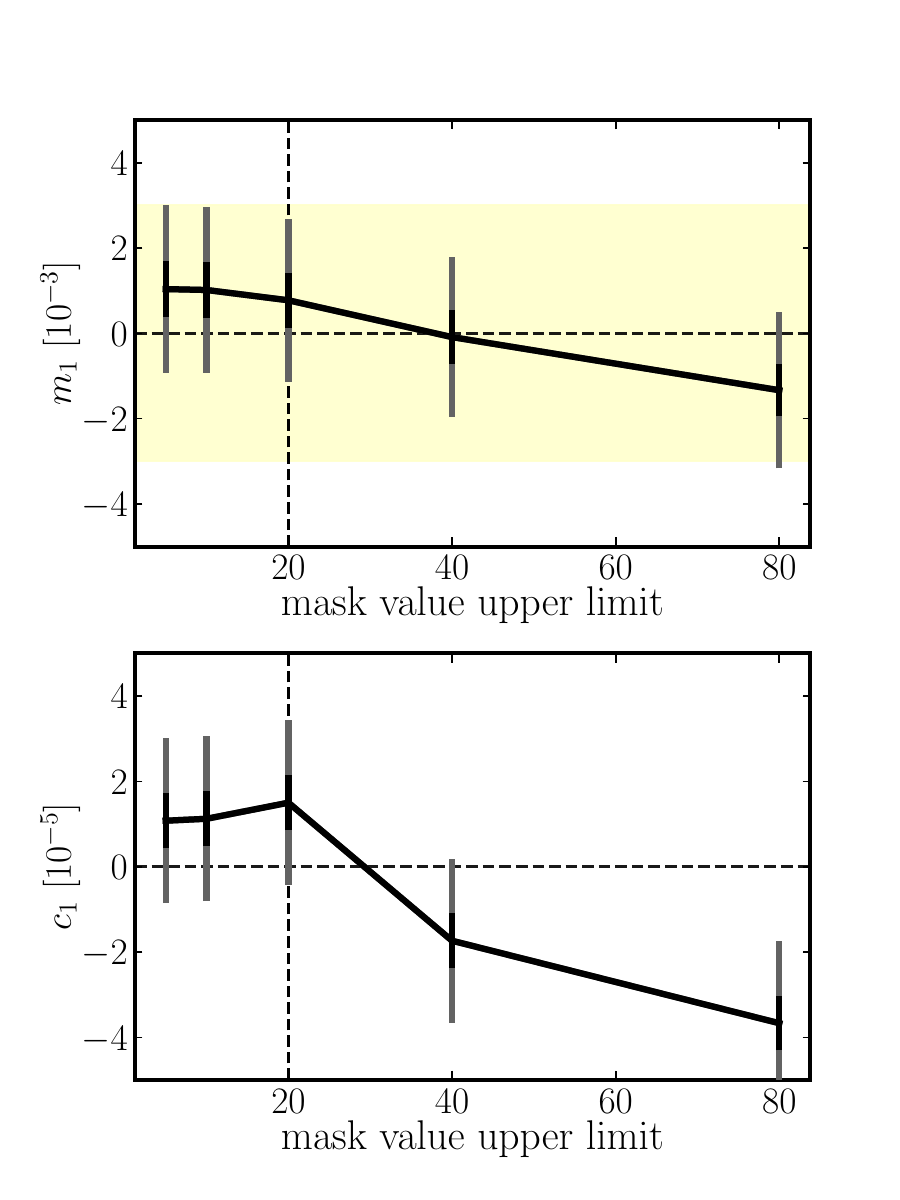}
\caption{
    Multiplicative bias (upper panel) and additive bias (lower panel) with
    1$\sigma$ and 3$\sigma$ errors as a function of the upper limit on the mask
    value. The vertical dashed line is the default upper limit on the mask
    value adopted in this paper to remove galaxies close to masked region. The
    shaded region are the LSST ten-year requirement on the control of
    systematic bias \citep{LSSTRequirement2018}.
    }
    \label{fig:test_mask}
\end{figure}

In Figure~\ref{fig:test_mask}, we illustrate the accuracy of the shear
estimator across different upper limits on the mask value. We find that when
increasing the upper limit to allow more sources near the masked regions, the
amplitude of multiplicative bias slightly increases where the amplitude of
additive bias significant increases. We find a significant negative additive
bias on the first component of shear if we do not remove galaxies close to the
mask using the mask value due to the vertical orientation of all bad columns
and bleed features in our simulation. This is because random camera rotation is
not applied  in the simulation. However, in the real LSST observations, single
visits at a particular spot have random orientations. The average additive bias
on sources that are close to the masked region should be smaller than what we
found here. Therefore, our selection and masking strategy is conservative for
the real LSST observations. Based on these findings, we recommend setting the
upper limit on the mask value to $20$. As depicted in
Figure~\ref{fig:test_mask}, this threshold ensures the multiplicative bias
meets LSST standards and the additive bias remains consistent with zero within
the statistical uncertainty of the test. Additionally, we report that the
additive systematic bias related to masking is below $4\times 10^{-5}$,
although the LSST DESC's requirement on the control of additive bias has not
been decided. The increased masked region for this threshold is demonstrated as
a grey region in the right panel of Figure~\ref{fig:test_msk_img_demo}.

\section{SUMMARY AND OUTLOOK}
\label{sec:summary}

This paper provides a comprehensive review and analytical proof of the noise
bias correction method for weak lensing shear inference, specifically focusing
on the ``renoising'' technique utilized within the \metacal{} framework. We
have integrated this analytical approach into the \anacal{} shear estimation
framework and introduced several significant enhancements to improve the
robustness and accuracy of the noise bias correction algorithm. To validate our
modified algorithm, we conducted simulations meticulously designed to emulate
the imaging characteristics of the LSST. These simulations included
semi-realistic galaxies and stars, characterized by distributions of magnitudes
and spatial densities that mirror those expected in actual observational data.

We rigorously tested our algorithm against a variety of challenging
observational conditions typical in high-precision astronomy. This included
images affected by cosmic rays, defective CCD columns, issues arising from
bright star saturation, bright star ``bleed trails'', and variations in the
point spread functions (PSFs) across the coadded image cell. The rigorous
testing regime was crafted to closely replicate the stringent conditions under
which the LSST operates assuming perfectly calibrated data --- perfect
astrometry, PSF modeling and masking.

The outcomes of our tests are promising, indicating that the multiplicative
shear bias in our algorithm is consistently less than 0.2 percent even under
many extreme conditions. The performance of our algorithm meets the demanding
requirement set by the LSST survey \citep{LSSTRequirement2018} even under many
extreme conditions, ensuring that it is well-suited for high-stakes
applications in cosmology where precision is paramount. This substantiates the
efficacy of our enhancements and the potential of our approach to contribute
significantly to the field of precision astronomy. Our algorithm achieves
exceptional computational speeds, processing each galaxy in less than a
millisecond. It can process all coadded images from the Rubin LSST using just a
few thousand CPU hours. The minimal computational resources required allow us
to conduct a large number of systematic tests on real observations even with
limited resources, ensuring that the shear measurements are sufficiently
accurate for weak lensing science. For instance, if we detect any additive
systematics during the null test that correlate with the observational
conditions of the survey, this would enable us to adjust the image processing
code. We could then rerun the entire sequence from the coadded image onward to
mitigate the identified systematics.

The \metadet{} algorithm has been validated in \citet{metaDet_LSST2023} and
will be applied to the LSST data. Since different shear estimators take
different assumptions, cross comparisons of any scientific results between the
catalogs from different shear estimators on the observed data are valuable.

This paper primarily focuses on evaluating the algorithm using constant-shear
simulations, in which all blended galaxies within a single image undergo
identical shear distortions. For simulations that incorporate
redshift-dependent shear, it may be necessary to simulate images that allow for
the joint calibration of both the redshift-dependent shear and the redshift
distribution of the source galaxies \citep{DESY3_BlendshearCalib_MacCrann2021,
zDependSim2023} The implications of redshift-dependent shear will be explored
in our future research.

\section*{ACKNOWLEDGEMENTS}
\addcontentsline{toc}{section}{ACKNOWLEDGEMENTS}

This paper has undergone internal review in the LSST Dark Energy Science
Collaboration. The primary authors would like to thank the internal reviewers
Arun Kannawadi and Gary Bernstein. We thank Erin Sheldon and Matthew Becker for
the help on using the package \texttt{descwl-shear-sims} and the comments on
the paper. In addition, we thank Erin Sheldon for comments on the paper draft.

This paper makes use of software developed for the Vera C.\ Rubin Observatory.
We thank the Vera C.\ Rubin Observatory for making their code available as free
software at http://dm.lsst.org.

We thank the maintainers of \numpy{} \citep{numpy_Harris2020}, \texttt{SciPy}
\citep{scipy_Virtanen2020}, \texttt{Matplotlib} \citep{matplotlib_Hunter2007},
\galsim{} \citep{GalSim} and \texttt{conda-forge} \citep{conda_forge} projects
for their excellent open-source software and software distribution systems.

Xiangchong Li and Rachel Mandelbaum are supported by the Department of Energy
grant DE-SC0010118. Xiangchong Li is an employee of Brookhaven Science
Associates, LLC under Contract No. DE-SC0012704 with the U.S. Department of
Energy.

The DESC acknowledges ongoing support from the Institut National de Physique
Nucléaire et de Physique des Particules in France; the Science \& Technology
Facilities Council in the United Kingdom; and the Department of Energy, the
National Science Foundation, and the LSST Corporation in the United States.
DESC uses resources of the IN2P3 Computing Center (CC-IN2P3–Lyon / Villeurbanne
- France) funded by the Centre National de la Recherche Scientifique; the
National Energy Research Scientific Computing Center, a DOE Office of Science
User Facility supported by the Office of Science of the U.S. Department of
Energy under Contract No. DE-AC02-05CH11231; STFC DiRAC HPC Facilities, funded
by UK BEIS National E-infrastructure capital grants; and the UK particle
physics grid, supported by the GridPP Collaboration. This work was performed in
part under DOE Contract DE-AC02-76SF00515.

Author contributions to this work are as follows: Xiangchong Li developed the
mathematical theory for the noise bias correction theory and designed the
pipeline for image simulation and shear estimation. Rachel Mandelbaum
supervised the project, providing valuable discussions and feedback on the
paper.

\section*{DATA AVAILABILITY}
\addcontentsline{toc}{section}{DATA AVAILABILITY}
The code used for this paper is publicly available on Github:
\url{https://github.com/mr-superonion/AnaCal/tree/v0.1.6}

\bibliographystyle{mnras}
\bibliography{citation}

\begin{thebibliography}{}
\makeatletter
\relax
\def\mn@urlcharsother{\let\do\@makeother \do\$\do\&\do\#\do\^\do\_\do\%\do\~}
\def\mn@doi{\begingroup\mn@urlcharsother \@ifnextchar [ {\mn@doi@} {\mn@doi@[]}}
\def\mn@doi@[#1]#2{\def\@tempa{#1}\ifx\@tempa\@empty \href {http://dx.doi.org/#2} {doi:#2}\else \href {http://dx.doi.org/#2} {#1}\fi \endgroup}
\def\mn@eprint#1#2{\mn@eprint@#1:#2::\@nil}
\def\mn@eprint@arXiv#1{\href {http://arxiv.org/abs/#1} {{\tt arXiv:#1}}}
\def\mn@eprint@dblp#1{\href {http://dblp.uni-trier.de/rec/bibtex/#1.xml} {dblp:#1}}
\def\mn@eprint@#1:#2:#3:#4\@nil{\def\@tempa {#1}\def\@tempb {#2}\def\@tempc {#3}\ifx \@tempc \@empty \let \@tempc \@tempb \let \@tempb \@tempa \fi \ifx \@tempb \@empty \def\@tempb {arXiv}\fi \@ifundefined {mn@eprint@\@tempb}{\@tempb:\@tempc}{\expandafter \expandafter \csname mn@eprint@\@tempb\endcsname \expandafter{\@tempc}}}

\bibitem[\protect\citeauthoryear{{Akeson} et~al.,}{{Akeson} et~al.}{2019}]{Roman2020}
{Akeson} R.,  et~al., 2019, \mn@doi [arXiv e-prints] {10.48550/arXiv.1902.05569}, \href {https://ui.adsabs.harvard.edu/abs/2019arXiv190205569A} {p. arXiv:1902.05569}

\bibitem[\protect\citeauthoryear{{Armstrong} et~al.,}{{Armstrong} et~al.}{2024}]{LSST_coaddCell2024}
{Armstrong} R.,  et~al., 2024, \mn@doi [arXiv e-prints] {10.48550/arXiv.2407.01771}, \href {https://ui.adsabs.harvard.edu/abs/2024arXiv240701771A} {p. arXiv:2407.01771}

\bibitem[\protect\citeauthoryear{Bartelmann \& Schneider}{Bartelmann \& Schneider}{2001}]{rev_wl_Bartelmann01}
Bartelmann M.,  Schneider P.,  2001, \mn@doi [Physics Reports] {https://doi.org/10.1016/S0370-1573(00)00082-X}, 340, 291

\bibitem[\protect\citeauthoryear{{Bernstein} \& {Armstrong}}{{Bernstein} \& {Armstrong}}{2014}]{BFD_Bernstein2014}
{Bernstein} G.~M.,  {Armstrong} R.,  2014, \mn@doi [\mnras] {10.1093/mnras/stt2326}, \href {http://adsabs.harvard.edu/abs/2014MNRAS.438.1880B} {438, 1880}

\bibitem[\protect\citeauthoryear{{Bernstein} \& {Jarvis}}{{Bernstein} \& {Jarvis}}{2002}]{Shapes_Bernstein2002}
{Bernstein} G.~M.,  {Jarvis} M.,  2002, \mn@doi [\aj] {10.1086/338085}, \href {http://adsabs.harvard.edu/abs/2002AJ....123..583B} {123, 583}

\bibitem[\protect\citeauthoryear{{Bernstein}, {Armstrong}, {Krawiec}  \& {March}}{{Bernstein} et~al.}{2016}]{BFD_Bernstein2016}
{Bernstein} G.~M.,  {Armstrong} R.,  {Krawiec} C.,   {March} M.~C.,  2016, \mn@doi [\mnras] {10.1093/mnras/stw879}, \href {http://adsabs.harvard.edu/abs/2016MNRAS.459.4467B} {459, 4467}

\bibitem[\protect\citeauthoryear{{Bosch} et~al.,}{{Bosch} et~al.}{2018}]{HSC1_pipeline}
{Bosch} J.,  et~al., 2018, \mn@doi [\pasj] {10.1093/pasj/psx080}, \href {http://adsabs.harvard.edu/abs/2018PASJ...70S...5B} {70, S5}

\bibitem[\protect\citeauthoryear{Bradbury et~al.,}{Bradbury et~al.}{2018}]{jax_Bradbury2018}
Bradbury J.,  et~al., 2018

\bibitem[\protect\citeauthoryear{{Chang} et~al.,}{{Chang} et~al.}{2013}]{WLsurvey_neffective_Chang2013}
{Chang} C.,  et~al., 2013, \mn@doi [\mnras] {10.1093/mnras/stt1156}, \href {https://ui.adsabs.harvard.edu/abs/2013MNRAS.434.2121C} {434, 2121}

\bibitem[\protect\citeauthoryear{{Harris} et~al.,}{{Harris} et~al.}{2020}]{numpy_Harris2020}
{Harris} C.~R.,  et~al., 2020, \mn@doi [\nat] {10.1038/s41586-020-2649-2}, \href {https://ui.adsabs.harvard.edu/abs/2020Natur.585..357H} {585, 357}

\bibitem[\protect\citeauthoryear{Hazimeh, Ponomareva, Mol, Tan  \& Mazumder}{Hazimeh et~al.}{2020}]{smoothstep_Hazimeh20}
Hazimeh H.,  Ponomareva N.,  Mol P.,  Tan Z.,   Mazumder R.,  2020, in III H.~D.,  Singh A.,  eds,  Proceedings of Machine Learning Research Vol. 119, Proceedings of the 37th International Conference on Machine Learning. PMLR, pp 4138--4148, \url {https://proceedings.mlr.press/v119/hazimeh20a.html}

\bibitem[\protect\citeauthoryear{{Heymans} et~al.,}{{Heymans} et~al.}{2006}]{Heymans2006}
{Heymans} C.,  et~al., 2006, \mn@doi [\mnras] {10.1111/j.1365-2966.2006.10198.x}, \href {https://ui.adsabs.harvard.edu/abs/2006MNRAS.368.1323H} {368, 1323}

\bibitem[\protect\citeauthoryear{{Heymans}, {Rowe}, {Hoekstra}, {Miller}, {Erben}, {Kitching}  \& {van Waerbeke}}{{Heymans} et~al.}{2012}]{PSFVar_Heymans2012}
{Heymans} C.,  {Rowe} B.,  {Hoekstra} H.,  {Miller} L.,  {Erben} T.,  {Kitching} T.,   {van Waerbeke} L.,  2012, \mn@doi [\mnras] {10.1111/j.1365-2966.2011.20312.x}, \href {https://ui.adsabs.harvard.edu/abs/2012MNRAS.421..381H} {421, 381}

\bibitem[\protect\citeauthoryear{{Huff} \& {Mandelbaum}}{{Huff} \& {Mandelbaum}}{2017}]{metacal_Huff2017}
{Huff} E.,  {Mandelbaum} R.,  2017, preprint, \href {http://adsabs.harvard.edu/abs/2017arXiv170202600H} {} (\mn@eprint {arXiv} {1702.02600})

\bibitem[\protect\citeauthoryear{{Hunter}}{{Hunter}}{2007}]{matplotlib_Hunter2007}
{Hunter} J.~D.,  2007, \mn@doi [Computing in Science and Engineering] {10.1109/MCSE.2007.55}, \href {https://ui.adsabs.harvard.edu/abs/2007CSE.....9...90H} {9, 90}

\bibitem[\protect\citeauthoryear{{Huterer}, {Takada}, {Bernstein}  \& {Jain}}{{Huterer} et~al.}{2006}]{shearSys_Huterer2006}
{Huterer} D.,  {Takada} M.,  {Bernstein} G.,   {Jain} B.,  2006, \mn@doi [\mnras] {10.1111/j.1365-2966.2005.09782.x}, \href {https://ui.adsabs.harvard.edu/abs/2006MNRAS.366..101H} {366, 101}

\bibitem[\protect\citeauthoryear{{Ivezi{\'c}} et~al.,}{{Ivezi{\'c}} et~al.}{2019}]{LSSTOverviwe2019}
{Ivezi{\'c}} {\v{Z}}.,  et~al., 2019, \mn@doi [\apj] {10.3847/1538-4357/ab042c}, \href {https://ui.adsabs.harvard.edu/abs/2019ApJ...873..111I} {873, 111}

\bibitem[\protect\citeauthoryear{{Kaiser}}{{Kaiser}}{2000}]{KSB_Kaiser2000}
{Kaiser} N.,  2000, \mn@doi [\apj] {10.1086/309041}, \href {https://ui.adsabs.harvard.edu/abs/2000ApJ...537..555K} {537, 555}

\bibitem[\protect\citeauthoryear{{Kilbinger}}{{Kilbinger}}{2015}]{rev_cosmicShear_Kilbinger15}
{Kilbinger} M.,  2015, \mn@doi [Reports on Progress in Physics] {10.1088/0034-4885/78/8/086901}, \href {http://adsabs.harvard.edu/abs/2015RPPh...78h6901K} {78, 086901}

\bibitem[\protect\citeauthoryear{{LSST Dark Energy Science Collaboration (LSST DESC)} et~al.,}{{LSST Dark Energy Science Collaboration (LSST DESC)} et~al.}{2021}]{DescDC2_2021}
{LSST Dark Energy Science Collaboration (LSST DESC)} et~al., 2021, \mn@doi [\apjs] {10.3847/1538-4365/abd62c}, \href {https://ui.adsabs.harvard.edu/abs/2021ApJS..253...31L} {253, 31}

\bibitem[\protect\citeauthoryear{{Laureijs} et~al.,}{{Laureijs} et~al.}{2011}]{Euclid2011}
{Laureijs} R.,  et~al., 2011, preprint, \href {http://adsabs.harvard.edu/abs/2011arXiv1110.3193L} {} (\mn@eprint {arXiv} {1110.3193})

\bibitem[\protect\citeauthoryear{{Li} \& {Mandelbaum}}{{Li} \& {Mandelbaum}}{2023}]{Anacal_Li2023}
{Li} X.,  {Mandelbaum} R.,  2023, \mn@doi [\mnras] {10.1093/mnras/stad890}, \href {https://ui.adsabs.harvard.edu/abs/2023MNRAS.521.4904L} {521, 4904}

\bibitem[\protect\citeauthoryear{{Li}, {Katayama}, {Oguri}  \& {More}}{{Li} et~al.}{2018}]{FPFS_Li2018}
{Li} X.,  {Katayama} N.,  {Oguri} M.,   {More} S.,  2018, \mn@doi [\mnras] {10.1093/mnras/sty2548}, \href {https://ui.adsabs.harvard.edu/abs/2018MNRAS.481.4445L} {481, 4445}

\bibitem[\protect\citeauthoryear{{Li} et~al.,}{{Li} et~al.}{2022a}]{HSC3_catalog}
{Li} X.,  et~al., 2022a, \mn@doi [\pasj] {10.1093/pasj/psac006}, \href {https://ui.adsabs.harvard.edu/abs/2022PASJ...74..421L} {74, 421}

\bibitem[\protect\citeauthoryear{{Li}, {Li}  \& {Massey}}{{Li} et~al.}{2022b}]{FPFS_Li2022}
{Li} X.,  {Li} Y.,   {Massey} R.,  2022b, \mn@doi [\mnras] {10.1093/mnras/stac342}, \href {https://ui.adsabs.harvard.edu/abs/2022MNRAS.511.4850L} {511, 4850}

\bibitem[\protect\citeauthoryear{{Li} et~al.,}{{Li} et~al.}{2023}]{zDependSim2023}
{Li} S.-S.,  et~al., 2023, \mn@doi [\aap] {10.1051/0004-6361/202245210}, \href {https://ui.adsabs.harvard.edu/abs/2023A&A...670A.100L} {670, A100}

\bibitem[\protect\citeauthoryear{{Li}, {Mandelbaum}, {Jarvis}, {Li}, {Park}  \& {Zhang}}{{Li} et~al.}{2024}]{Anacal_Li2024}
{Li} X.,  {Mandelbaum} R.,  {Jarvis} M.,  {Li} Y.,  {Park} A.,   {Zhang} T.,  2024, \mn@doi [\mnras] {10.1093/mnras/stad3895}, \href {https://ui.adsabs.harvard.edu/abs/2024MNRAS.52710388L} {527, 10388}

\bibitem[\protect\citeauthoryear{{MacCrann} et~al.,}{{MacCrann} et~al.}{2022}]{DESY3_BlendshearCalib_MacCrann2021}
{MacCrann} N.,  et~al., 2022, \mn@doi [\mnras] {10.1093/mnras/stab2870}, \href {https://ui.adsabs.harvard.edu/abs/2022MNRAS.509.3371M} {509, 3371}

\bibitem[\protect\citeauthoryear{{Mandelbaum}}{{Mandelbaum}}{2018}]{rev_wlsys_Mandelbaum2017}
{Mandelbaum} R.,  2018, \mn@doi [\araa] {10.1146/annurev-astro-081817-051928}, \href {https://ui.adsabs.harvard.edu/abs/2018ARA\&A..56..393M} {56, 393}

\bibitem[\protect\citeauthoryear{{Mandelbaum} et~al.,}{{Mandelbaum} et~al.}{2018a}]{HSC1_GREAT3Sim}
{Mandelbaum} R.,  et~al., 2018a, \mn@doi [\mnras] {10.1093/mnras/sty2420}, \href {https://ui.adsabs.harvard.edu/abs/2018MNRAS.481.3170M} {481, 3170}

\bibitem[\protect\citeauthoryear{{Mandelbaum} et~al.,}{{Mandelbaum} et~al.}{2018b}]{HSC1_shapecalib}
{Mandelbaum} R.,  et~al., 2018b, \mn@doi [\mnras] {10.1093/mnras/sty2420}, \href {https://ui.adsabs.harvard.edu/abs/2018MNRAS.481.3170M} {481, 3170}

\bibitem[\protect\citeauthoryear{{Massey} \& {Refregier}}{{Massey} \& {Refregier}}{2005}]{polar_shapelets_Massey2005}
{Massey} R.,  {Refregier} A.,  2005, \mn@doi [\mnras] {10.1111/j.1365-2966.2005.09453.x}, \href {http://adsabs.harvard.edu/abs/2005MNRAS.363..197M} {363, 197}

\bibitem[\protect\citeauthoryear{{Massey} et~al.,}{{Massey} et~al.}{2007}]{galsim_STEP2}
{Massey} R.,  et~al., 2007, \mn@doi [\mnras] {10.1111/j.1365-2966.2006.11315.x}, \href {https://ui.adsabs.harvard.edu/abs/2007MNRAS.376...13M} {376, 13}

\bibitem[\protect\citeauthoryear{{Massey} et~al.,}{{Massey} et~al.}{2013}]{WLsys_Massey2013}
{Massey} R.,  et~al., 2013, \mn@doi [\mnras] {10.1093/mnras/sts371}, \href {https://ui.adsabs.harvard.edu/abs/2013MNRAS.429..661M} {429, 661}

\bibitem[\protect\citeauthoryear{{Moffat}}{{Moffat}}{1969}]{Moffat1969}
{Moffat} A.~F.~J.,  1969, \aap, \href {http://adsabs.harvard.edu/abs/1969A\%26A.....3..455M} {3, 455}

\bibitem[\protect\citeauthoryear{{Pujol}, {Kilbinger}, {Sureau}  \& {Bobin}}{{Pujol} et~al.}{2019}]{preciseSim_Pujol2019}
{Pujol} A.,  {Kilbinger} M.,  {Sureau} F.,   {Bobin} J.,  2019, \mn@doi [\aap] {10.1051/0004-6361/201833740}, \href {https://ui.adsabs.harvard.edu/abs/2019A\&A...621A...2P} {621, A2}

\bibitem[\protect\citeauthoryear{{Refregier}}{{Refregier}}{2003}]{shapeletsI_Refregier2003}
{Refregier} A.,  2003, \mn@doi [\mnras] {10.1046/j.1365-8711.2003.05901.x}, \href {http://adsabs.harvard.edu/abs/2003MNRAS.338...35R} {338, 35}

\bibitem[\protect\citeauthoryear{{Refregier}, {Kacprzak}, {Amara}, {Bridle}  \& {Rowe}}{{Refregier} et~al.}{2012}]{noiseBiasRefregier2012}
{Refregier} A.,  {Kacprzak} T.,  {Amara} A.,  {Bridle} S.,   {Rowe} B.,  2012, \mn@doi [\mnras] {10.1111/j.1365-2966.2012.21483.x}, \href {http://adsabs.harvard.edu/abs/2012MNRAS.425.1951R} {425, 1951}

\bibitem[\protect\citeauthoryear{{Rowe} et~al.,}{{Rowe} et~al.}{2015}]{GalSim}
{Rowe} B.~T.~P.,  et~al., 2015, \mn@doi [Astronomy and Computing] {10.1016/j.ascom.2015.02.002}, \href {http://adsabs.harvard.edu/abs/2015A\%26C....10..121R} {10, 121}

\bibitem[\protect\citeauthoryear{{Sanchez}, {Mendoza}, {Kirkby}, {Burchat}  \& {LSST Dark Energy Science Collaboration}}{{Sanchez} et~al.}{2021}]{weaklensDeblend}
{Sanchez} J.,  {Mendoza} I.,  {Kirkby} D.~P.,  {Burchat} P.~R.,   {LSST Dark Energy Science Collaboration} 2021, \mn@doi [\jcap] {10.1088/1475-7516/2021/07/043}, \href {https://ui.adsabs.harvard.edu/abs/2021JCAP...07..043S} {2021, 043}

\bibitem[\protect\citeauthoryear{{Sheldon} \& {Huff}}{{Sheldon} \& {Huff}}{2017}]{metacal_Sheldon2017}
{Sheldon} E.~S.,  {Huff} E.~M.,  2017, \mn@doi [\apj] {10.3847/1538-4357/aa704b}, \href {http://adsabs.harvard.edu/abs/2017ApJ...841...24S} {841, 24}

\bibitem[\protect\citeauthoryear{{Sheldon}, {Becker}, {MacCrann}  \& {Jarvis}}{{Sheldon} et~al.}{2020}]{metaDet_Sheldon2020}
{Sheldon} E.~S.,  {Becker} M.~R.,  {MacCrann} N.,   {Jarvis} M.,  2020, \mn@doi [\apj] {10.3847/1538-4357/abb595}, \href {https://ui.adsabs.harvard.edu/abs/2020ApJ...902..138S} {902, 138}

\bibitem[\protect\citeauthoryear{{Sheldon}, {Becker}, {Jarvis}, {Armstrong}  \& {LSST Dark Energy Science Collaboration}}{{Sheldon} et~al.}{2023}]{metaDet_LSST2023}
{Sheldon} E.~S.,  {Becker} M.~R.,  {Jarvis} M.,  {Armstrong} R.,   {LSST Dark Energy Science Collaboration} 2023, \mn@doi [The Open Journal of Astrophysics] {10.21105/astro.2303.03947}, \href {https://ui.adsabs.harvard.edu/abs/2023OJAp....6E..17S} {6, 17}

\bibitem[\protect\citeauthoryear{{Spergel} et~al.,}{{Spergel} et~al.}{2015}]{Roman_Spergel2015}
{Spergel} D.,  et~al., 2015, preprint, \href {http://adsabs.harvard.edu/abs/2015arXiv150303757S} {} (\mn@eprint {arXiv} {1503.03757})

\bibitem[\protect\citeauthoryear{{The LSST Dark Energy Science Collaboration} et~al.,}{{The LSST Dark Energy Science Collaboration} et~al.}{2018}]{LSSTRequirement2018}
{The LSST Dark Energy Science Collaboration} et~al., 2018, arXiv e-prints, \href {https://ui.adsabs.harvard.edu/abs/2018arXiv180901669T} {p. arXiv:1809.01669}

\bibitem[\protect\citeauthoryear{{Virtanen} et~al.,}{{Virtanen} et~al.}{2020}]{scipy_Virtanen2020}
{Virtanen} P.,  et~al., 2020, \mn@doi [Nature Methods] {10.1038/s41592-019-0686-2}, \href {https://ui.adsabs.harvard.edu/abs/2020NatMe..17..261V} {17, 261}

\bibitem[\protect\citeauthoryear{Zhang}{Zhang}{2008}]{Z08}
Zhang J.,  2008, \mn@doi [\mnras] {10.1111/j.1365-2966.2007.12585.x}, 383, 113

\bibitem[\protect\citeauthoryear{{Zhang}, {Sheldon}  \& {Becker}}{{Zhang} et~al.}{2023}]{deepfield_metacal2023}
{Zhang} Z.,  {Sheldon} E.~S.,   {Becker} M.~R.,  2023, \mn@doi [The Open Journal of Astrophysics] {10.21105/astro.2206.07683}, \href {https://ui.adsabs.harvard.edu/abs/2023OJAp....6E..16Z} {6, 16}

\bibitem[\protect\citeauthoryear{conda-forge community}{conda-forge community}{2021}]{conda_forge}
conda-forge community 2021, {The conda-forge Project: Community-based Software Distribution Built on the conda Package Format and Ecosystem}, \mn@doi{10.5281/zenodo.4774217}, \url {https://doi.org/10.5281/zenodo.4774217}

\bibitem[\protect\citeauthoryear{{von K{\'a}rm{\'a}n}}{{von K{\'a}rm{\'a}n}}{1948}]{PSFVar_vonK1948}
{von K{\'a}rm{\'a}n} T.,  1948, \mn@doi [Proceedings of the National Academy of Science] {10.1073/pnas.34.11.530}, \href {https://ui.adsabs.harvard.edu/abs/1948PNAS...34..530V} {34, 530}

\makeatother
\end{thebibliography}
\appendix

\section{Pipeline}
\label{app:pipeline}
We briefly summarize the pipeline for ensemble shear estimation as follows:
\begin{enumerate}
    \item The center of every pixel is considered as a candidate of galaxy.
    \item Apply a binary pixel mask to the image to set regions near bright
        stars to zero.
    \item Add a pure noise image with the same statistical properties, but
        rotated by 90°, to the image.
    \item
        Deconvolve the PSF and convolve the image with a Gaussian to calculate
        the peak detection modes ($\dtilde{q}_i$) for each pixel and compute
        the detection weight ($\dtilde{w}_d$).
    \item Select the peak pixels with $\dtilde{w}_d > 0$ as peaks of detected
        sources, and use them as origins for the measurements below.
    \item Create a continuous smoothed version of the binary pixel mask from
        step (i) and eliminate detections at pixels that fall above a threshold
        in this smoothed mask.
    \item Measure the shapelet modes ($\dtilde{M}_{nm}$), their shear
        responses, and the shear responses of the peak detection modes;
    \item Compute the nonlinear observables, including selection weight
        ($\dtilde{w}_s$) and ellipticity ($\dtilde{\epsilon}_{1,2}$);
    \item Select the source with $w_s>0$ to compute the weighted ellipticity
        $e_{1,2}$\,.
    \item Measure the measurement errors of shapelet modes ($\delta M_{nm}'$)
        and the peak detection modes ($\delta q_i'$) and their shear responses
        from the pure noise image at the position of the detected peaks.
    \item Compute the shear responses of the weighted ellipticity following
        equation~\eqref{eq:shear_response_final}.
    \item Estimate shear with equation~\eqref{eq:shear_estimator_renoise}.
\end{enumerate}

\section{Noise Response}
\label{app:noise_bias}

In Section~\ref{sec:method_noise_renoise}, we demonstrate our ability to add an
additional layer of noise, identical in statistics to the original noise after
rotating by $90^\circ$ counterclockwise, to an image and subsequently derive the
linear shear response from the galaxy image with two layers of noise. Since the
shear distortion affects only the galaxy and not the noise, the estimated shear
response is calculated by subtracting the effective shear responses of the two
noise images from the overall shear response of the noisy image.

In this appendix, we validate the claim made in
equation~\eqref{eq:noise_response} that the expected effective shear responses
of each noise image are identical:
\begin{equation}
\begin{split}
&\left\langle
    \left.
    \partial_i e_1\!(\vnu + \delta \vnu + \delta \vnu' )
    \right|_{\vec{\gamma} = 0}
    \delta \nu'_{;1i}
\right\rangle\\
&\qquad=
    \left\langle
    \left.
    \partial_i e_1\!(\vnu + \delta \vnu + \delta \vnu' )
    \right|_{\vec{\gamma} = 0}
    \delta \nu_{;1i}
    \right\rangle\,,
\end{split}
\end{equation}
where $\partial_i e_1$ refers to the $i$th element of the gradient vector of
$e_1$\,. $\delta \vnu$ is the linear observable vector measured from the
original noise image. $\delta \vnu'$ is the linear observable vector measured
from the additional noise image.

We start from the left-hand side:
\begin{equation}
\left\langle
    \left.
    \partial_i e_1\!(\vnu + \delta \vnu + \delta \vnu' )
    \right|_{\vec{\gamma} = 0}
    \delta \nu'_{;1i}
\right\rangle\,,
\end{equation}
which remains unchanged if we rotate the galaxies and the two levels of image
noises all together by $90\degr$. To prove equation~\eqref{eq:noise_response},
we rotate the galaxies by $90\degr$ clockwise and recompute the left-hand side.
This is equivalent to rotating both the original image noise and the additional
image noise by $90\degr$ counterclockwise, while keeping the galaxies
unchanged. Since the shapes of galaxies are isotropically oriented and their
positions are randomly distributed, the estimation from the rotated galaxy
sample remains the same. Consequently, we have
\begin{equation}
\begin{split}
&
\left\langle
    \left.
    \partial_i e_1\!(\vnu + \delta \vnu + \delta \vnu' )
    \right|_{\vec{\gamma} = 0}
    \delta \nu'_{;1i}
\right\rangle\\
& \qquad =
\left\langle
    \left.
    \partial_i e_1\!(\vnu + \delta \vnu' + \delta \vnu'' )
    \right|_{\vec{\gamma} = 0}
    \delta \nu''_{;1i}
\right\rangle\,,
\end{split}
\end{equation}
where $\delta \vnu''$ has the same statistics with the original noise after
being rotated by $180\degr$ counterclockwise. To be more specific, the rotation
transforms $\delta \vnu$ to $\delta \vnu'$ and transforms $\delta \vnu'$ to
$\delta \vnu''$\,. We note that for homogeneous noise, rotating the coordinate
by $180\degr$ does not change the statistics of the noise field; therefore, we
can replace $\delta \vnu''$ with $\delta \vnu$ in the last equation and obtain
equation~\eqref{eq:noise_response}.

\label{lastpage}
\end{document}